\documentclass[lettersize,journal]{IEEEtran}
\usepackage[table]{xcolor}% http://ctan.org/pkg/xcolor
\usepackage{cite}
\usepackage{amsmath,amssymb,amsfonts}
\newcommand{\ignore}[1]{ }
\usepackage{algorithmic}
\usepackage{graphicx}
\usepackage{textcomp}
\usepackage{xcolor}
\usepackage{float}
\usepackage{multirow}
\usepackage{lipsum}
\usepackage{pifont}

\usepackage{array}

\usepackage{balance}

\usepackage{hyperref}

\hypersetup{
  colorlinks,
  allcolors=blue
  citebordercolor=blue,
  citecolor=blue,
  linkcolor=blue,
  urlcolor=blue}

\usepackage{courier}

\usepackage{ragged2e}

\newcommand{\red}{}

\newcommand{\blue}{}
\renewcommand{\red}{\textcolor{red}}

\renewcommand{\blue}{\textcolor{blue}}

\usepackage[prependcaption, textsize=footnotesize]{todonotes}
\newcommand{\hassan}[1]{\todo[inline,linecolor=black,backgroundcolor=orange!30,bordercolor=red]{Hassan: #1}}

\newcommand{\Amir}[1]{\todo[inline,linecolor=black,backgroundcolor=green!30,bordercolor=blue]{Amir: #1}}

\usepackage{tikz,xcolor,hyperref}

\definecolor{lime}{HTML}{A6CE39}
\DeclareRobustCommand{\orcidicon}{%
	\begin{tikzpicture}
	\draw[lime, fill=lime] (0,0) 
	circle [radius=0.16] 
	node[white] {{\fontfamily{qag}\selectfont \tiny ID}};
	\draw[white, fill=white] (-0.0625,0.095) 
	circle [radius=0.007];
	\end{tikzpicture}
	\hspace{-2mm}
}

\usepackage{mathtools}
\DeclarePairedDelimiter\ceil{\lceil}{\rceil}

\foreach \x in {A, ..., Z}{%
	\expandafter\xdef\csname orcid\x\endcsname{\noexpand\href{https://orcid.org/\csname orcidauthor\x\endcsname}{\noexpand\orcidicon}}
}

\begin{document}

% "Sorting It Out in Hardware: A Comparative Study
% "Hardware-Assisted Sorting: A State-of-the-Art Survey"
%"Hardware-Assisted Sorting: A Comprehensive Review and Comparison"
% "Sorting in the Fast Lane: A Survey of Hardware-Enhanced Sorting"
% "Sorting in the Fast Lane: Hardware-Driven Sorting Approaches"
%Hardware-Powered Sorting: A Survey of Techniques and Trends
%"Sorting It Out: A State-of-the-Art Survey of Hardware-Accelerated Sorting Methods"
%Harnessing Hardware for Sorting: An Extensive Survey of Methods
%Sorting Faster with Hardware: A Survey of Accelerated Sorting Techniques

\title{\huge Sorting it out in Hardware: A State-of-the-Art Survey}
%\title{Hardware-Powered Sorting:\\ A State-of-the-Art Survey}

%\title{Hardware-Assisted Sorting: \red{A Survey on Literature and Future Considerations}}

%Review of Hardware Sorting Techniques
%should not be use}

\author{Amir Hossein Jalilvand \orcidA{}, Faeze~S.~Banitaba \orcidB{},  Seyedeh~Newsha~Estiri \orcidC{},\\ Sercan Aygun~\orcidD{}, \textit{Member, IEEE}, and M.~Hassan~Najafi \orcidE{}, \textit{Senior Member, IEEE}

%\vspace{-1em}

\thanks{The authors are with the School of Computing and Informatics,
University of Louisiana, LA, 70504, USA. Email: \{amir.jalilvand, faeze-sadat.banitaba1, seyedeh-newsha.estiri1, sercan.aygun, najafi\}@louisiana.edu This work was supported in part by National Science Foundation (NSF) grant \#2019511, the Louisiana Board of Regents Support Fund \#LEQSF(2020-23)-RD-A-26, and generous gifts from Cisco, Xilinx, and NVIDIA.}
}

\maketitle

\begin{abstract}

Sorting is a fundamental operation in various applications and %has been 
a traditional research topic in computer science.  
%The improvement in 
Improving the performance of sorting operations %algorithms 
can have a significant impact on many application domains. % plays an important role in essential role in various ranges of applications.
%To date
For high-performance sorting, much attention has been paid to hardware-based solutions. These %solutions 
are often realized with application-specific integrated circuits (ASICs) or field-programmable gate arrays (FPGAs). Recently, in-memory sorting solutions have also been proposed to address the movement cost issue between memory and processing units, also known as Von Neumann bottleneck.
%Because of 
Due to the complexity of the sorting algorithms, achieving an efficient hardware implementation for sorting data is challenging. %of sorting algorithms is a challenging problem. 
%Most of the proposed 
A large body of prior solutions is built %based 
on compare-and-swap (CAS) units. These %solutions are called
are categorized as \textit{comparison-based} sorting. Some recent solutions offer 
%Recently, some solutions have been proposed taking advantage of removing the compare-and-swap units and are called 
\textit{comparison-free} sorting. %solutions.  
In this survey, we review
%In this survey review paper, we survey 
the latest works in the area of hardware-based sorting. %of hardware solutions for sorting. 
We also discuss the recent hardware solutions for \textit{partial} and \textit{stream} sorting. %and stream sorting. %, %which is 
%a relaxed version of the sorting problem. 
%\amir
%\red{
%In-memory Computation %(IMC)-- %or CIM: Computing-in-Memory or  -- %depending on the computing granularity) 
%is a promising solution %paradigm for solving 
%to address this Von Neumann bottleneck.
%Recently, in-memory sorting solutions have been also proposed. 
%We seek to cover these works in this survey.
Finally, we will discuss %raise 
some important concerns %challenges 
that need to be considered %addressed 
in the future designs of sorting systems. %proposals.}

\end{abstract}
%\vspace{1em}

%\vspace{-1em}
\section{Introduction}
\begin{figure}[t!]
    %\vspace{-1.7em}
	\centering
	\includegraphics[width=160pt]{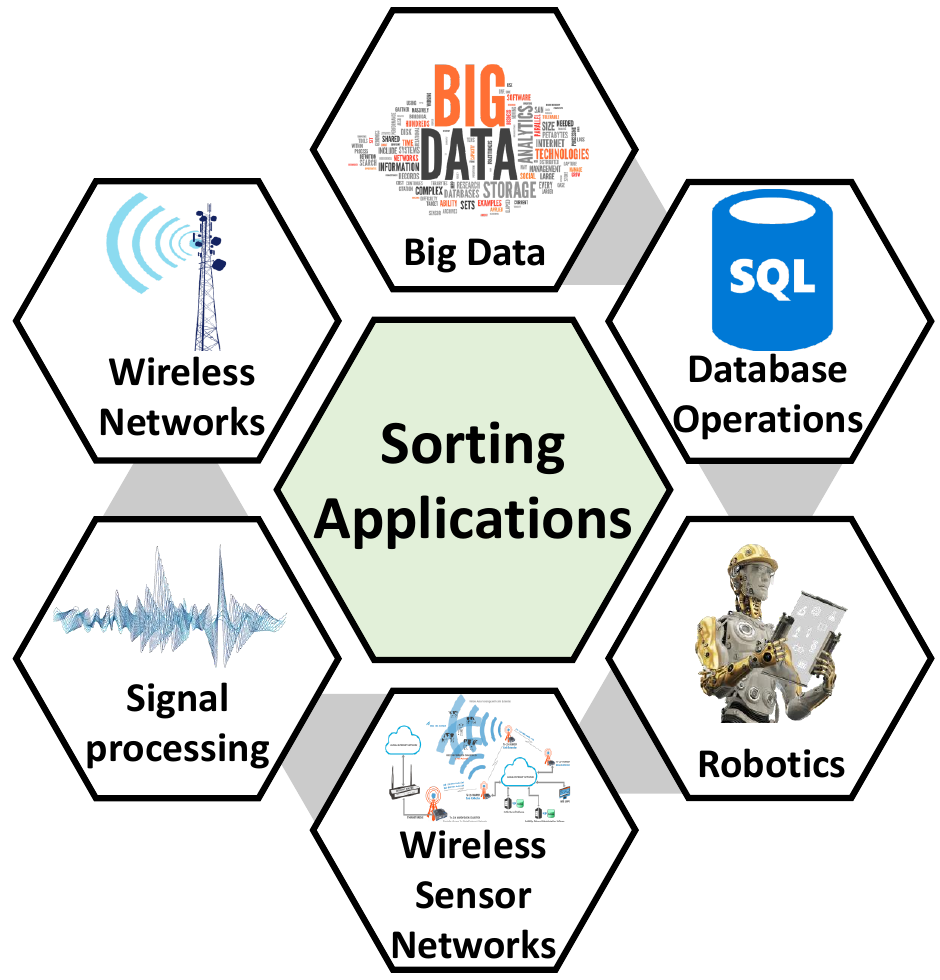}
%\vspace{-1.em}
	\caption{{Common applications of sorting:
 \textit{Big Data} \cite{MUNRO1980315, 10.1145/3373087.3375304, 6114577},
 \textit{Database operations} \cite{10.1007/978-3-642-69419-6_12, 10.1145/1132960.1132964, 10.1145/253262.253322}, \textit{Robotics} \cite{7095088, DBLP:journals/corr/abs-2009-03565, GUO2023104271, Shaikat_Akter_Salma_2020}, \textit{Wireless Sensor Networks} \cite{electronics11193024, bordimSorting, 1192726, https://doi.org/10.1049/wss2.12048, 9585092}, \textit{Signal Processing} \cite{1041030, 6147654, 10.1007/3-540-44968-X_25, 9292936, 8261869}, and \textit{Wireless Networks} \cite{Shiau2006GeneralizationOS}.}
 }
\label{fig:appliation}
	%\vspace{-1em}
\end{figure}

Today, the data volume 
has increased significantly in many application domains. Processing data at the terabyte and petabyte levels has become routine. %challenging. %reality. 
Processing large volumes of data is challenging and %this volume of data 
%has faced a severe challenge and %this increase 
is expected to remain at an upward rate~\cite{Fan2013Mining}. 
Sorting is one of the %most 
substantial operations in computer science 
%The purpose of sorting is to put a list of data in a specific order (\textit{Ascending} or \textit{Descending}). 
%\sercan{}
%\red{Or: 
%Sorting is done 
performed for different purposes, from %The purposes of sorting can be listed as: (i) 
putting data in a specific order,  such as
\textit{ascending} or \textit{descending}, %(ii) 
to find the minimum and maximum values, %in a set of data, %the 
 %(iii) 
finding %calculating 
the median, and %(iv) 
partial sorting to find %getting 
the top-$m$ greatest or smallest values. %}
As %shown in 
Fig.~\ref{fig:appliation} shows sorting is used in many application domains, %operation 
%some operations like 
%data merging, and %it 
%performance is crucial in a wide range of applications, %including 
from data merging to big data processing~\cite{mankowitz2023faster, pang2022application}, database operations~\cite{do2023efficient} especially when the scale of files/data are very large, robotics~\cite{DEZAKI202260, Montesdeoca2022Analysis, shirvani2023threshold}, signal processing (\textit{e.g.,} sorting radar signals)~\cite{9388166, 1041030, 6147654}, and wireless networks~\cite{8589084}.
%\Amir{I added the following line}
%\sercan{I agree, thank you.}

Sorting the time series data according to their timestamps holds critical importance in numerous artificial intelligence (AI) applications, such as forecasting and anomaly detection \cite{anomaly_det}, where the sequential occurrence of events is of paramount significance~\cite{10184865}.
Wireless sensor network applications often incorporate genetic algorithms, with the `Non-dominated Sorting Genetic Algorithm (NSGA)' being a commonly employed and efficient approach requiring sorting~\cite{996017}. Additionally, wireless networks necessitate the implementation of sorting algorithms that are both energy-optimal and energy-balanced, such as enhanced sorting algorithms \cite{1192726}. The concept of sorting also extends to the realm of robotic visual tasks. Much like traditional scalar sorting, the sorting of items based on attributes like color, shape, or other features within a robot's perceived environment constitutes a tangible engineering application of sorting~\cite{Shaikat_Akter_Salma_2020}. In the field of robotics, object sorting is a significant task. Particularly in computer vision applications, sorting objects by robots based on their perceived environment is challenging~\cite{DBLP:journals/corr/abs-2009-03565}.
%When robots attempt to sort objects based on their perceived environment, particularly in computer vision applications, it presents a challenging endeavor~\cite{DBLP:journals/corr/abs-2009-03565}. 
Another intriguing application %draws parallels between a specific domain and a hardware-aware sorting approach, which is leveraged 
is to control greenhouse climatic factors through sorting networks~\cite{doi:10.1177/1550147718756871}. For sorting large-scale datasets, some researchers adopt an external sorting methodology. External sorting serves as a solution for sorting vast datasets that cannot fit into the primary memory of a computing platform. Instead, it utilizes additional memory elements like hard disk drives, employing a sort-and-merge strategy~\cite{Chen_2020}. %Furthermore, 
Sorting also finds unconventional applications in signal processing. %the context of signals and their processing. 
This extends from theoretical scalar sorting to sorting tasks in real-world signal processing. %and sorting tasks. 
An illustrative example is radar signal sorting, %which represents 
a recent and intricate sorting challenge in the context of multi-function radar systems~\cite{9292936}. %As emerging hardware platforms continue to evolve, recent technology introduces new memory environments. The concept of `approximate memory' describes memory systems that make a trade-off between perfect data fidelity and increased storage density, as well as reduced energy consumption for both memory and input/output operation; approximate memory sorting applications occur as an emerging area \cite{10.1145/3299874.3318036}}

\begin{table*}
\caption{Comparison between the existing surveys for hardware solutions of sorting\\ {(\ding{109}: not covered, \ding{119}: partially covered, \ding{108}: fully covered )}}
%\vspace{-0.5em}
\label{tab:survey}
\centering
\begin{tabular}{|cc|c|c|c|c|c|}
\hline
\multicolumn{1}{|c|}{Article} &  Year & \begin{tabular}[c]{@{}c@{}}Sorting \\ Networks\end{tabular} & \begin{tabular}[c]{@{}c@{}}Comparison-based\\  Sorting\end{tabular} & \begin{tabular}[c]{@{}c@{}}Comparison-free\\  Sorting\end{tabular} & Partial Sorting & In-memory Sorting\\ \hline
\multicolumn{1}{|c|}{Jmaa \textit{et al.}~\cite{ben2019comparative}}   & 2019 & \ding{109}                                                         & \ding{108}                                                                 & \ding{109}                                                                & \ding{109}                                                          & \ding{109}                                                             \\ \hline
\multicolumn{1}{|c|}{Skliarova~\cite{electronics11071029}}   & 2021 & \ding{108}                                                         & \ding{108}                                                                 & \ding{109}                                                                   & \ding{109}                                                          & \ding{109}                                                             \\ \hline
\multicolumn{1}{|c|}{Ali~\cite{ali2022hardware}}   & 2022 & \ding{119}                                                         & \ding{119}                                                                 &\ding{119}                                                                & \ding{109}                                                          & \ding{109}                                                            \\ \hline
\multicolumn{2}{|c|}{\textbf{This work}}  & \ding{108}                                                         & \ding{108}                                                                 & \ding{108}                                                               & \ding{108}                                                       & \ding{108}                                                          \\ \hline
\end{tabular}
\vspace{-1em}
\end{table*}

%Fig.~\ref{fig:appliation} shows the most common applications of sorting. 
Improving the sorting speed %performance
%of data sorting %algorithms 
can have a significant impact on all %many of 
these %above-mentioned
%The enhancement in the performance of sorting algorithms plays an essential role in 
applications. %mentioned above. %Lots of 
Many software- and hardware-based  solutions have been proposed in the literature for high-performance sorting. %so far. 
Software-based solutions rely on %include 
powerful single/multi-core and graphics processing unit (GPU)-based processors for high performance~\cite{Singh2018Survey}. Much attention has been paid to hardware sorting solutions, especially for  applications that require very high-speed sorting \cite{Farmahini2013Modular, Najafi2018Low-Cost, Montesdeoca2022Analysis}. These %solutions are 
have been implemented using either application-specific integrated circuits (ASICs) or field-programmable gate arrays (FPGAs). %Based 
Depending on the target applications, the hardware sorting units vary greatly in how they are configured and implemented. The number of inputs can be as low as nine for some image processing applications (\textit{e.g.,} median filtering~\cite{Draz2023}) %or as high as 
to tens of thousands~\cite{Najafi2018Low-Cost, Farmahini2013Modular}. The data inputs have been %are sometimes 
binary values, integers, or floating-point numbers ranging from 4- to 256-bit precision. 

Hardware cost and power consumption are the dominant concerns with hardware implementations. The total chip area is limited in many applications~\cite{10.14778/1454159.1454171}. As fabrication technologies continue to scale, keeping chip temperatures low is an important goal since leakage current increases exponentially with temperature. Power consumption must be kept as low as possible. Developing low-cost, power-efficient hardware-based solutions to sorting is an important goal~\cite{Najafi2018Low-Cost}.

There is a large body of work %in the literature 
on the design of customized sorting hardware.  %for sorting algorithms. 
These works seek %are proposed 
to utilize the hardware resources fully and to provide a custom, cost-effective hardware sorting engine. %processing. 
Developing hardware-efficient %hardware 
implementations for sorting algorithms is challenging, %\red{quarrelsome} \sercan{maybe cumbersome? or challenging?} 
%taking into account 
considering the %inherent
complexity of these algorithms~\cite{Abdel2017An, Farmahini2013Modular, Najafi2018Low-Cost}. A significant amount of hardware resources is spent by %considerable quantity of hardware resources is necessitated, made up of %, but not limited to,
comparators, memory elements including large global memories, %and 
complex pipelining, and %in addition to 
complicated local and global control units~\cite{Abdel2017An}. %Most of the proposed 

\begin{figure}[t!]
    %\vspace{-1.5em}
	\centering
	\includegraphics[trim={0.5cm 0.8cm 0.5cm 0.5cm},clip,width=0.55\linewidth]{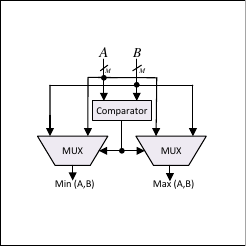}
	%\vspace{-1.5em}
	\caption{Compare-and-Swap (CAS) operation in hardware. }
	\label{fig1}
	\vspace{-1em}
\end{figure}

Many of the prior hardware solutions are built on basic %based on
compare-and-swap (CAS) units that compare pairs of data and swap if needed. %which consist of a 
These solutions are categorized as %called 
\textit{comparison-based} sorting. %since they employ the CAS units. 
As shown in Fig.~\ref{fig1}, each basic CAS unit is conventionally implemented with a binary comparator and two multiplexers (\texttt{MUX}) units~\cite{Najafi2018Low-Cost}. %However, due  to the  continual  CAS operations, these solutions suffer from latency and resource consumption issues, especially for large databases. 
Sorting networks of CAS units are frequently used for fast and parallel hardware sorting. %solution for hardware-based sorting. 
Their inherent parallelism enables them 
to achieve sorting at a considerably faster rate than the fastest sequential software-based sorting algorithms. However, these CAS-based hardware solutions suffer from high hardware costs, especially when the \textit{number} and \textit{precision} of input data increase~\cite{Ray2022K-Degree}.
%In the last five years, some comparison-free/quasi-comparison-free sorting solutions have been proposed taking advantage of the removal/reduction of CAS blocks.
In the last few years, some \textit{comparison-free/quasi-comparison-free} sorting solutions have been proposed to address the challenges with \textit{comparison-based} sorting designs. %by removing the CAS blocks. 
We will discuss these novel solutions in Section~\ref{Sec:Compfree}.

\begin{table*}[t]
\centering
\caption{Prior art for hardware solutions of sorting}
\label{tab:survey2}
%\vspace{-0.5em}
\begin{tabular}{|c|l|c|l|c|}
\hline
Category                           & Reference                                                                & Year & \multicolumn{1}{c|}{Idea}                                                                                   & Design\\ \hline
\multirow{7}{*}{\begin{tabular}[c]{@{}c@{}}\textbf{Comparison} \\ \textbf{Based}\end{tabular}} \multirow{8}{*}{}  & Farmahini \textit{et al.}\cite{Farmahini2013Modular} & 2013 & Sorting   using hierarchical smaller blocks      & \ding{171}                     \\ \cline{2-5} 
                                   & Lin   \textit{et al.}\cite{lin2017hardware}                     & 2017 & Pointer-like   iterative architecture                            & \ding{171}
                          %    \\ \cline{2-5} 
%& +++++                    & 2017 & Pointer-like   iterative architecture                                                                       & check             
\\ \cline{2-5}                           
                                   & Najafi   \textit{et al.}\cite{Najafi2018Low-Cost}               & 2018 & Bit-stream-based and time-encoded unary design                             & \ding{171}        \\ \cline{2-5} 
                                   & Norollah   \textit{et al.}\cite{norollah2019rths}               & 2019 & Consecutive   normal and reverse sorting                   & \ding{67}                              \\ \cline{2-5} 
                                   & Jelodari   \textit{et al.}\cite{JELODARI2020103203}             & 2020 & Vertex   indexing in graph representation of inputs         & \ding{67}              \\ \cline{2-5} 
    &{Papaphilippou \textit{et al}~\cite{Papaphilippou2020AN, papaphilippou2018flims}}             & 2020 & {Recursive parallel merge tree}       & {\ding{67}}          \\ \cline{2-5} 
    & Preethi   \textit{et al.}\cite{Preethi2021L}             & 2021 & {Clock gating techniques to improve  power consumption}      & \ding{171}   \\ 
    %\ignore{\cline{2-5}     & {Axtmann   \textit{et al.}\cite{axtmann2022engineering} }            & 2022 & {{A Versatile Multi-core Algorithm}}      & {\ding{75}---}   \\}   
    \cline{2-5} 
                                   & {Prince \textit{et al.}\cite{prince2023scalable} }            & 2023 & {Sorting weighted stochastic bit-streams}      & 
 {\ding{171}}   \\       \hline

\multirow{6}{*}{\begin{tabular}[c]{@{}c@{}}\textbf{Comparison} \\ \textbf{Free}\end{tabular}
} \multirow{7}{*}{}  & Abdel-Hafeez \textit{et al.} \cite{Abdel2017An}      & 2017 & One-hot   weight representation                                                                             & \ding{171}
                              \\ \cline{2-5} 
                                   & Bhargav \textit{et al.} \cite{bhargav2019power}           & 2019 & FSM   module for sorting                                                                                    & \ding{171}
                              \\ \cline{2-5} 
                                   & Chen   \textit{et al.}\cite{Chen2021A}                          & 2021 & Bidirectional architecture improving sorting cycle \cite{bhargav2019power} & \ding{171}
/ \ding{67}                         \\ \cline{2-5} 
                                   & Sri   \textit{et al.}\cite{Sri2022An}                           & 2022 & Improving   the boundary-finding module of \cite{Chen2021A}                                & \ding{67}                              \\ \cline{2-5} 
                                   & Ray \textit{et al.} \cite{Ray2022K-Degree}                 & 2022 & $k$-Degree   Parallel Comparison-free Hardware Sorter                                                         & \ding{67}                              \\ \cline{2-5} 
                                   & Jalilvand   \textit{et al.}\cite{Amir2022A}                     & 2022 & Comparison-free   sorting architecture based on unary computing                                             & \ding{171}
                              \\ \hline
\multirow{8}{*}{\begin{tabular}[c]{@{}c@{}}\textbf{Partial} \\ \textbf{Sorting}\end{tabular}
} \multirow{9}{*}{}  & Yu   \textit{et al.} \cite{6132381}                            & 2011 & Spike   sorting hardware                                                                                    & \ding{67}                              \\ \cline{2-5} 
                                   & Campobello   \textit{et al.} \cite{CAMPOBELLO2012178}           & 2012 & Maximum   - minimum finder                                                                                  & \ding{67}                              \\ \cline{2-5} 
                                   & Subramaniam   \textit{et al.} \cite{SubramaniamMedian}          & 2017 & Median   finder                                                                                             & \ding{67}                              \\ \cline{2-5} 
                                   & Korat   \textit{et al.} \cite{8249010}                          & 2017 & Odd-even   merge sorting                                                                                    & \ding{67}                              \\ \cline{2-5} 

                                   & Valencia \textit{et al.} \cite{spikingMIN}           & 2019 & Minimum   distance calculation for spike sorting                                                            & \ding{67}                              \\ \cline{2-5} 
                                   & Zhang   \textit{et al.} \cite{AngiziPIMmax}                     & 2021 & Min-max   sorting architecture                                                                              & \ding{87}                 \\ \cline{2-5} 
                                   & Yan   \textit{et al.} \cite{minMaxFPGA}                         & 2021 & Determining   a certain $k$ largest or smallest numbers                                                     & \ding{67}                              \\ \hline
\multirow{10}{*}{\begin{tabular}[c]{@{}c@{}}\textbf{In-} \\ \textbf{Memory}\end{tabular}} %& 
%Kvatinsky \textit{et al.} \cite{kvatinsky2014magic}        & 2014 &         \textit{Memristor-Aided Logic}.                                         &  \ding{71}                  \\ \cline{2-5} 
%& Trancoso \textit{et al.} \cite{trancoso2015moving}         & 2015 & Moving to memoryland  & \ding{70}                     \\ \cline{2-5} 
    & {Wu \textit{et al.}\cite{wu2015data}  }                                   & {2015} & {Data sorting in flash memory (NAND flash-based)} & {\ding{118}} \\ \cline{2-5} 
        %& {Laga \textit{et al.}  \cite{7932112}  }                                 & {2017} & {MONTRAS-NVM hybrid memory-aware sorting algorithm} & {\ding{74} / \ding{70}}  \\ \cline{2-5} 
    & {Samardzic \textit{et al.} \cite{samardzic2020bonsai} }    & {2020} & {Bonsai Sorting on CPU-FPGA focused on DRAM-scale sorting} & {\ding{67} / \ding{118}}  \\ \cline{2-5} 
                                   & Li \textit{et al.} \cite{li2020imc}                        & 2020 & Parallel sorting via hybrid memory cubes                                                      & \ding{115}                 \\ \cline{2-5} 
                                   & Prasad \textit{et al.} \cite{prasad2021memristive}      & 2021 & %\red{Memresistor}
                                   Memristor-based data ranking and min/max computation                             & \ding{70}                     \\ \cline{2-5} 
                                   %& \red{ \textit{et al.} \cite{}  }      & \red{2021} & hardware   co-optimization for in-memory sorting                                                            & \ding{75}                 \\ \cline{2-5} 
        & Chu \textit{et al.} \cite{chu2021nvmsorting}               & 2021 & \ignore{NVM sorting: }Detecting partially ordered for cost reduction & \ding{74}     \\ \cline{2-5} 
                                   & Riahi Alam \textit{et al.} \cite{alam2022sorting}              & 2022 & High-performance and energy-efficient data sorting     & \ding{72}                  \\ \cline{2-5} 
                                   & Yu \textit{et al.} \cite{yu2022fast}                                       & 2022 & Column-skipping algorithm for\ignore{different banks of }memristive memory & \ding{168}    \\ \cline{2-5} 
                                   & {Zokaee \textit{et al.}\cite{zokaee2022sky}}   & {2022} & Sorting large datasets based on sample sort\ignore{Sky-Sorter: In-memory architecture for large-scale sorting} & {\ding{70}}  \\ \cline{2-5} 
                                   & Lenjani \textit{et al.} \cite{lenjani2022pulley, lenjani2020fulcrum}                  & 2022 & Optimizing external sorting for NVM-DRAM hybrid storage\ignore{$n$-memory-layer approach for multi-gigabyte sorting}                                                       & \ding{70}           \\ \cline{2-5} 
                                   & {Liu \textit{et al.} \cite{liu2023lazysort}  }                                  & {2023} & Minimizing NVM write operations\ignore{Lazysort: A NVM-DRAM hybrid memory} & {\ding{118}}  \\ \hline
\end{tabular}
\justify{\scriptsize{\ding{171} $\xrightarrow{}$ ASIC, \ding{67} $\xrightarrow{}$ FPGA, \ding{87} $\xrightarrow{}$ DRAM, \ding{70} $\xrightarrow{}$ In-Memory, \ding{168} $\xrightarrow{}$ Memristor-based, \ding{72} $\xrightarrow{}$ Adaptive Memristor, \ding{73} $\xrightarrow{}$ In-Logic-Layer Based, \ding{74} $\xrightarrow{}$ NVM-Based, \linebreak \ignore{\ding{75} $\xrightarrow{}$ \red{In-Memory-Layer Accelerator}, }\ding{115} $\xrightarrow{}$ In-Logic-Layer Based, \ding{118} $\xrightarrow{}$ In-Memory Friendly}}
\vspace{-0.5em}
\end{table*}

%\sercan{
\textit{Complete} sorting %methods
sorts all items ($N$) of a list. 
\textit{Partial} sorting has also been a popular sorting variant. %is another variant of the sorting. 
Unlike complete sorting, %which %aims to 
%sorts all items ($N$) of a list, %a list of items ($n$) in an order, 
partial sorting returns a list of the $k$ smallest or largest elements in order where $K < N$~\cite{partialSortInitialPaper, Farmahini2013Modular}. 
The cost of partial sorting is often substantially less than complete sorting 
%The virtue of partial sorting %, where possible,
%is that the cost is substantially less than for sorting 
when the number of sub-sorting attempts 
%\red{the number of order statistics} \sercan{I guess this is "the number of sub-sorting attempts"} 
%\hassan{Or it should be K? For example for N=128 I guess we have many subsorting computations even for partial sorting.}
%required 
is small compared to $N$.
The other elements (above the $k$ smallest ones) may also be sorted as in-place partial sorting or discarded, which is common in streaming partial sorts \cite{chambers1971Algorithm}.
%With the increase of the amount of data, 

%A common practical example of partial sorting is computing the "Top 100" of some list.
%\Amir{In this paper}

Despite many recent %interesting 
works in hardware-assisted sorting, 
%Regardless of the existence of \red{(Considering the existence of)} new ideas (\red{studies}) in hardware-assisted sorting, 
no recent survey %paper 
%exists to 
reviews the latest developments in this area. % in this field (\red{an extensive survey to fully cover the most recent works in this field is needed}).
%Reviewing 
Studying the literature, we found three 
%Our research finding confirms that there are three recent 
surveys discussing prior %in the field of 
hardware-based sorting designs. These are compared in % as shown in
Table~\ref{tab:survey}. Jmaa \textit{et al.}~\cite{ben2019comparative} compare the performance of the hardware implementations of popular sorting algorithms (\textit{i.e.}, \texttt{Bubble Sort}, \texttt{Insertion Sort}, \texttt{Selection Sort}, \texttt{Quick Sort}, \texttt{Heap Sort}, \texttt{Shell Sort}, \texttt{Merge Sort}, and \texttt{Tim Sort}) in terms of execution time, standard deviation, and resource utilization. They synthesized the designs %were synthesized in the
on a Zynq-7000 FPGA platform. Skliarova~\cite{electronics11071029} reviewed different implementation approaches for network-based hardware accelerators for \textit{sorting}, \textit{searching}, and \textit{counting} tasks. %networks.
Ali~\cite{ali2022hardware} looked closely at comparison-based and comparison-free hardware solutions for sorting. %The mentioned survey papers did not consider partial sorting and in-memory sorting solutions at all. (\red{or: 
As in-memory 
%sorting
%computing 
and partial sorting are relatively emerging topics, % in the %categories of
%sorting literature, 
these previous surveys do not cover them.  %completely.
%}) 
%Motivated by this, in this survey paper, we review  the latest  hardware sorting solutions in (\red{I guess giving them in a sentence is better; that way, we stress the list less (if we will give a list, maybe we can reconsider the list), and I am not quite sure beginning with in-memory is a good idea or not. So, maybe:}). \red{
Motivated by this, %we 
this work reviews  the latest hardware %sorting 
solutions %considering 
for complete, partial and in-memory sorting, covering both comparison-based and comparison-free approaches. %We look in the literature for the latest %at the 
%comparison-based and comparison-free approaches to  %complete 
%sorting.
%literature.
%}
%\Amir{We may change this category}
%two main categories:
% \begin{itemize}\item Complete sorting,      \item Partial sorting
%\end{itemize}
%We also review in-memory solutions in each category. (\red{this maybe removed}) 
%In addition, 
Table~\ref{tab:survey2} summarizes and classifies the important works we study in this %review 
article.

The remainder of this paper is organized as follows.  
Section~\ref{Sec:CS} reviews complete sorting solutions. %will be reviewed.
Section~\ref{Sec:PS} reviews hardware solutions for partial sorting.
Section~\ref{Sec:InMemSorting} discusses recent works on emerging in-memory sorting.
Section~\ref{Sec:Ochall} discusses open challenges and future works. Finally, Section~\ref{Sec:Conclusion} concludes the paper.

%\vspace{-0.75em}
\section{Complete Sorting Methods}
\label{Sec:CS}

We begin by reviewing %In this section, we review 
recent works on complete sorting, which  %survey on the complete sorting methods.
%Complete sorting 
processes %and evaluates
all the data to sort them %be sorted 
in an ascending or descending order. We divide our discussion into two categories of \textit{comparison-based} and \textit{comparison-free} sorting.%} 

\subsection{Comparison-based}

%\sercan{I am adding an intro here as it immediately appears with literature} 

\begin{figure*}
    %\vspace{-1em}
\centering
	\includegraphics[trim={0cm 3.5cm 0cm 0.2cm},clip,width=0.75\linewidth]{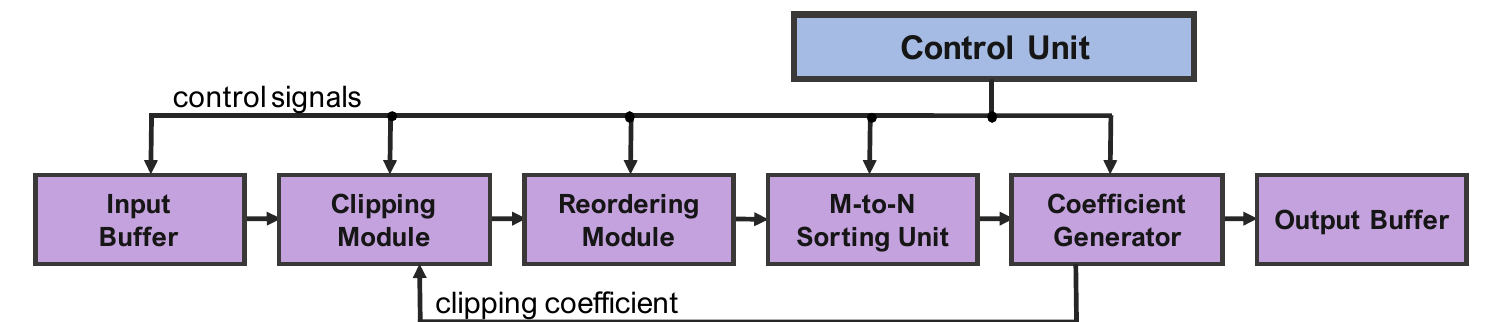}
	%\vspace{-.5em}
	\caption{Lin \textit{et al.}~\cite{lin2017hardware}
 %The proposed low-power 
 iterative sorting system. %'s architecture is presented. 
 An iterative architecture is designed by repeatedly employing a smaller sorting unit to process streaming input data. Within this iterative max-set-selection or partial sorting system, the input remains constant, contingent on the type of sorting unit in use. Users have the flexibility to select different sorting units as the Core Module, allowing them to strike a balance between throughput and resource constraints. Importantly, this iterative architecture imposes no limitations on the volume of input samples it can handle. As the input size scales, resource consumption remains constant, effectively mitigating resource overhead challenges. In addition to the application of the low-power sorting module, the design also incorporates an adaptive clipping mechanism and a reordering module. These elements are instrumental in further reducing the switching activities of registers. The adaptive clipping coefficient increases in pair with the temporal results during the sorting process, serving to block a substantial number of samples.}
	\label{fig-iterative}
	\vspace{-1em}
%\Amir{I suggest to have a long caption similar to Fig.15 }
%\Faeze{done}
\end{figure*}

%In 2013, 
Farmahini \textit{et al.}~\cite{Farmahini2013Modular} proposed a %notable 
comparison-based design that employs efficient techniques for constructing high-throughput, low-latency sorting units using smaller building blocks in a hierarchical manner. Their design includes %the 
$N$-to-$M$ \textit{sorting} and \textit{max-set-selection} units. %which are %both 
%adaptable, %and offer 
%low-latency and high-throughput. %Additionally, 
They extensively discuss the structure, performance, and resource requirements of these units. %, making their approach a valuable contribution to the field of sorting network design. 
Despite its primary focus on integer numbers, their design %the design proposed in this study 
%exhibits the capability to 
efficiently accommodates two's complement and floating-point numbers, %without difficulty, 
as the comparators utilized in their compare-and-exchange (CAE) blocks can be substituted accordingly.
%\ignore{\hassan{What ``CAE'' stands for here?}}

Some sorting applications do not need to sort all input data. 
%In some sorting applications, %various scenarios, 
%there may not be a need to sort all input values. 
Instead, %certain applications 
the application may only require the identification of the $M$ largest or $M$ smallest values from a set of $N$ inputs. %Here, M is less than N, and both are integers that are powers of two, \textit{i.e.}, M equals 2 to the power of m and N equals 2 to the power of n, where m and n are integers. 
These algorithms are called \textit{partial sorters} and will be discussed 
later in this survey. %Being so, 
In an $N$-to-$M$ max-set-selection unit used in the sorting designs of~\cite{Farmahini2013Modular}, only the $M$ largest inputs are required in no specific order. %without any specific ordering. 

%In 2017, 
Lin \textit{et al.}~\cite{lin2017hardware} proposed a hardware acceleration architecture for %the 
real-time sorting of $M$ %N 
%values 
out of $N$ %M 
inputs. %in a large data set.
Their design benefits from moving indexes instead of data %itself; 
%They call it 
and is called a \textit{pointer-like} design. They %try to 
reduce power consumption %gain low-power consumption 
by reducing switching activities and signal transitions while %having 
maintaining high throughput. Their sorting approach has a %They claim to have 
complexity of %equal to
$O(\log_{2}^2 M)$. The primary contributor to power consumption is the switching activities of registers. To effectively reduce power, they recommend modification to the register transfer level (RTL) design. Notably, signal transitions increase when the input dataset is larger or when the bit width of the input sample is significant. They propose to incorporate additional registers to represent the position of each input sample. So, only the indexes need to be migrated from register to register. When $N$ inputs are present, the complete index can be represented using only $\log_{2} N$ bits, irrespective of whether the bit-widths are 8-bit, 16-bit, or more. %wider. 
While modifications may increase the total cell area, they achieve a substantial reduction in %substantially reduce 
dynamic power dissipation.

Executing the sorting process using a single module is impractical for large input datasets, as it requires high %higher 
I/O bandwidth and large %a larger total 
cell area. %particularly for large %when %with 
%increasing 
%input size. 
To %overcome 
mitigate this issue, %they 
Lin \textit{et al.}~\cite{lin2017hardware} proposed %their solution is 
to reuse smaller sorting units as the core module and combine these small units with other control units to implement an iterative architecture. Fig.~\ref{fig-iterative} shows their proposed architecture.  %The architecture of their %proposed 
%sorting system is shown %depicted 
 Users have the flexibility to select different sorting units as the core module, enabling them to trade off throughput for resource constraints.

\begin{figure}
    %\vspace{-1.7em}
	\centering
\includegraphics[trim={0.5cm 1.0cm  0.5cm 0.7cm}, clip, width=0.5\linewidth]{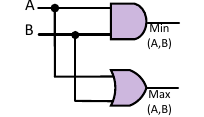}
	%\vspace{-1.5em}
	\caption{The hardware implementation of a  Unary \textit{CAS} block%.% (a) Conventional binary design. (b) Unary design
 ~\cite{Najafi2018Low-Cost}.}
 %\Amir{We have the binary CAS block in Fig2. I reformed this figure}
	\label{fig4}
	\vspace{-1em}
\end{figure}

%In 2018, 
%another group of researchers, 
Najafi \textit{et al.}~\cite{Najafi_Sortin_ICCD'17, Najafi2018Low-Cost} developed %presented 
an area- and power-efficient hardware design for complete sorting based on \textit{unary} computing (UC). They convert the data from binary to unary bit-streams to sort them in the unary domain. Their approach replaces the conventional complex design of the \textit{CAS} unit implemented based on binary radix with a simple unary-based design made of simple standard \texttt{AND} and \texttt{OR} gates.
Fig.~\ref{fig4} demonstrates how a \textit{CAS} block is 
implemented in %conventional binary and 
the unary domain.
When two unary bit-streams of equal length are connected to the inputs, an \texttt{AND} gate yields the minimum value, whereas an \texttt{OR} gate produces the maximum value. %When correlated stochastic bit-streams are used as inputs, these gates demonstrate analogous functionality~\cite{alaghi2013stochastic}.
An overhead of this unary %their proposed %hardware 
design is the cost of converting data from binary to unary representation. However, compared to the cost savings in the computation circuits, this conversion overhead is insignificant. They report an area and power saving of more than 90\% for implementing a 256-input complete sorting network. %processing algorithm in which they added a circuit to convert data between binary and unary format and they claim this overhead is small;
%however, Najafi \textit{et al.} emphasize that 
The unary design of~\cite{Najafi2018Low-Cost} %Their proposed logic 
%is made from 
consists of simple logic gates %while being 
independent of data size. %As we discussed earlier, the main part of a \textit{CAS} block is the maximum and minimum calculation, which is performed with simple hardware in the unary domain. %: an \texttt{OR} gate for the maximum and an \texttt{AND} gate for the minimum operation. % and one \texttt{OR} gate implement a unary \textit{CAS} block. 
The computation %precision and 
accuracy is controlled by the length of bit-streams. The longer the bit-stream, the higher the accuracy.
But processing %The inherent data structure of 
long unary bit-streams can %potentially 
result in long latency with the sorting design of~\cite{Najafi2018Low-Cost}. This %Processing long unary bit-streams often 
causes runtime overhead compared to the conventional binary process. %so that the stream size should be discussed effectively suitable to the application.%\ignore{\red{I am not surely using they're here. this is informal. Also, a kinder way to say this drawback is better like: "; however, Najafi \textit{et al.} underscore that the inherent data structure of long streams may cause latency}\blue{done}\red{processing long may cause runtime overhead, so that the stream size should be discussed effectively suitable to the application}\blue{done}} 
%Therefore, Najafi \textit{et al.} underscore that the runtime overhead is tolerable in many applications. 
While the latency may be tolerated in many applications, they introduce a time-based unary design to mitigate the latency issue. They encode the input data %using 
to pulse-width modulated signals. The data value is determined by the duty cycle in this approach. %is encoded in time and then represented by pulse signals.
%So, they were able to gain improvements in latency and energy consumption with a slight decrease in accuracy. 
At the cost of slight accuracy loss, the time-based approach significantly reduces the latency. %and energy consumption. 

\ignore{
\red{this part may be more detailed like that \texttt{AND} gates and \texttt{OR} gates do here - min and max selection details.}\blue{done}}

\begin{figure}[t]
    %\vspace{-1.7em}
	\centering
	\includegraphics[trim={0cm 0cm 0cm 0cm},clip,width=1.0\linewidth]{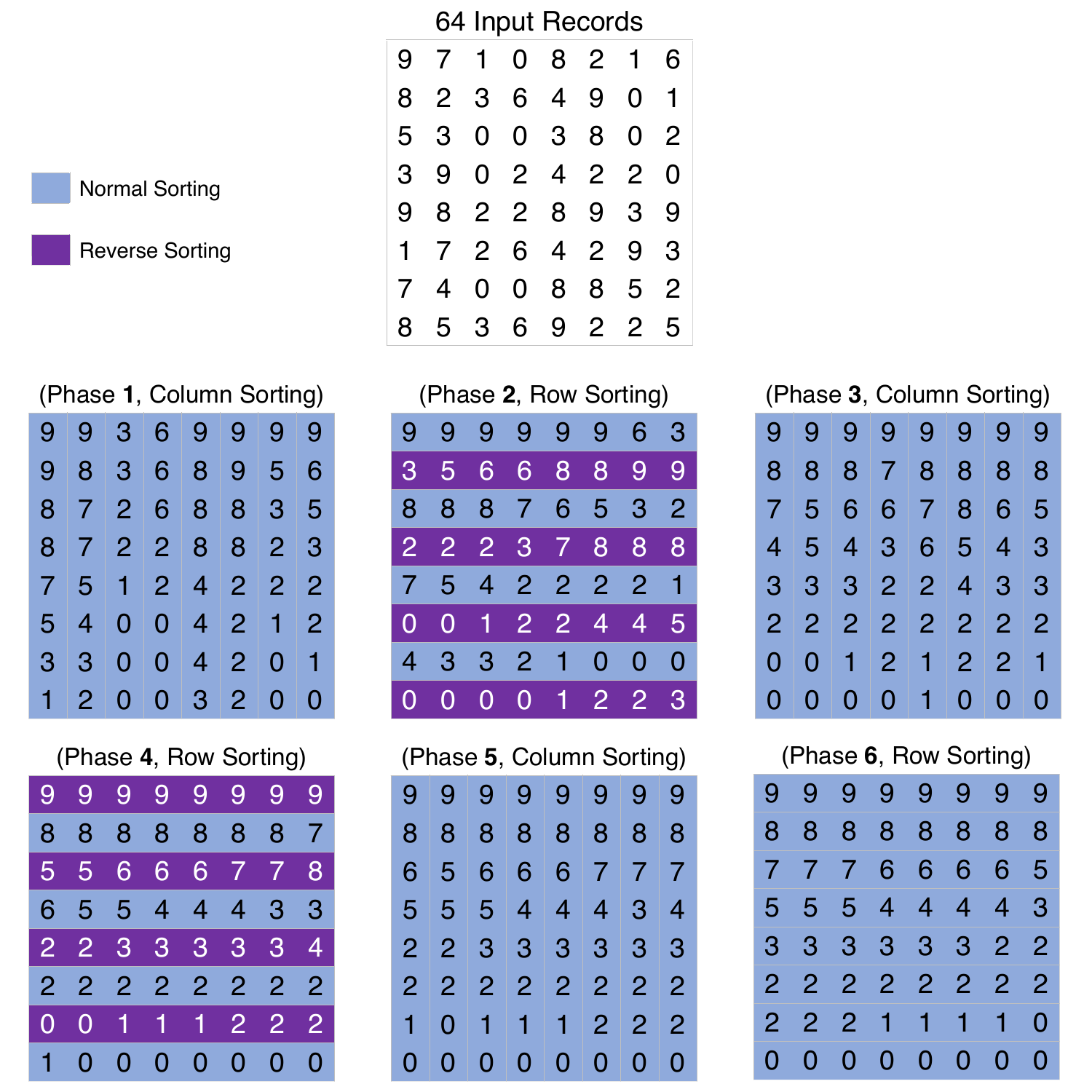}
	%\vspace{-2.em}
	\caption{The MDSA %in action 
 with 64 input records, forming an 8×8 matrix\cite{norollah2019rths}.}
	\label{fig8}
	%\vspace{-1.5em}
\end{figure}

Prince \textit{et al.}~\cite{prince2023scalable} combined the bit-stream capabilities of stochastic computing (SC) with binary weighting, reducing latency of bit-stream-based sorting. The approach offers good scalability and cost-efficiency compared to SC and traditional binary methods, making it an efficient solution for sorting tasks. They use a weighted bit-stream converter to generate weighted bit-streams for an adaptable sorting network. Unlike conventional SC bit-streams, each bit in the weighted bit-streams retains its weight as a standard binary value. This conversion reduces the number of bits in SC from $2^N$ to $N$ for $N$-bit precision, resulting in a substantial reduction in latency and energy consumption by shifting from exponential to linear representation. 
They propose a new lock-and-swap (LAS) unit to sort weighted bit-streams. Their LAS-based sorting network can determine the result of comparing different input values early and then map the inputs to the corresponding outputs based on shorter weighted bit-streams.
%The determination of a comparison between two bit-streams occurs as soon as the initial disparity between the two input streams is detected.

%Later in 2019, 
Norollah \textit{et al.}~\cite{norollah2019rths} presented a novel multidimensional sorting algorithm (MDSA) and its corresponding architecture, a real-time hardware sorter (RTHS), to efficiently sort large sequences of records. MDSA %The proposed algorithm significantly 
reduces the required resources, enhances memory efficiency, and has a minimal negative impact on execution time, even when the number of input records increases. To sort large sequences of records, MDSA %the proposed algorithm 
divides a sequence into smaller segments, which are then sorted separately. As shown in Fig.~\ref{fig8}, the MDSA algorithm consists of six consecutive phases and two modes: normal and reverse sorting. %the scenario wherein 
The sorting network organizes the records in descending and ascending order for normal and reverse modes, respectively. %Then 
In each phase, the sorting networks are fed by a group of input records to sort independently.
%with each sorting network assigned some records to sort independently at each phase.
%\hassan{This needs to be revised.}\Faeze{done}
%As shown in Fig.~\ref{fig8}, at each phase, 

\begin{figure}
    %\vspace{-1.7em}
	\centering
	\includegraphics[trim={0cm 0cm 0cm 0.9cm},clip,width=.9\linewidth]{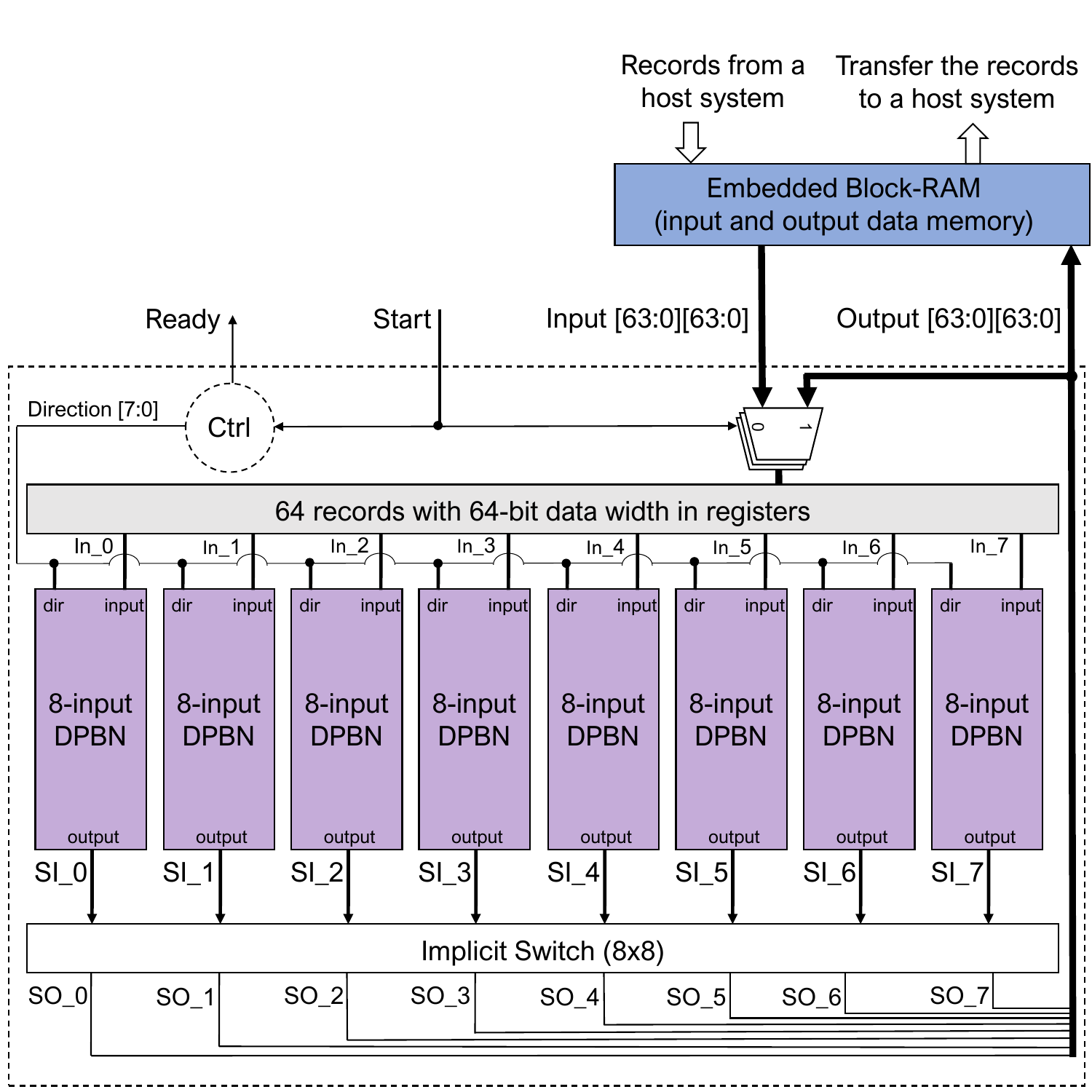}
	%\vspace{-1.em}
	\caption{Real-time hardware sorter (RTHS) architecture for 8×8 matrix records~\cite{norollah2019rths}.}
	\label{fig5}
	%\vspace{-0.5em}
\end{figure}

The authors in~\cite{norollah2019rths} claim that their sorting method is more beneficial for resource conservation (memory efficiency) while providing high performance. Fig.~\ref{fig5} shows the complete architecture of RTHS. %In this design, 
In this design, pipelining is used %as a technique 
to reduce the critical path in dual-mode pipeline bitonic networks (DPBNs). Fig.~\ref{fig6} shows a DPBN unit for 8 inputs. %This is illustrated in Fig.~\ref{fig6}. 
The number of pipeline stages in a DPBN is directly proportional to its number of steps, which can be computed by $(1/2 \log_{2}(N)(\log_{2}(N) + 1))$, where $N$ is the number of inputs. The implicit switch is done by fixed wiring and so is completely static. This hardwired switching %\red{implemented ? we may remove this} component 
does not require additional routing resources and has minimal overhead.
%\Faeze{by hardwired component, I'm referring to the implicit switch, that was mentioned in previous sentence. is it better to say "this fixed wiring switch"?}

\ignore{

\begin{figure}
    %\vspace{-1.7em}
	\centering
	\includegraphics[trim={0.5cm 0.5cm 0.5cm 0.5cm},clip,width=0.9\linewidth]{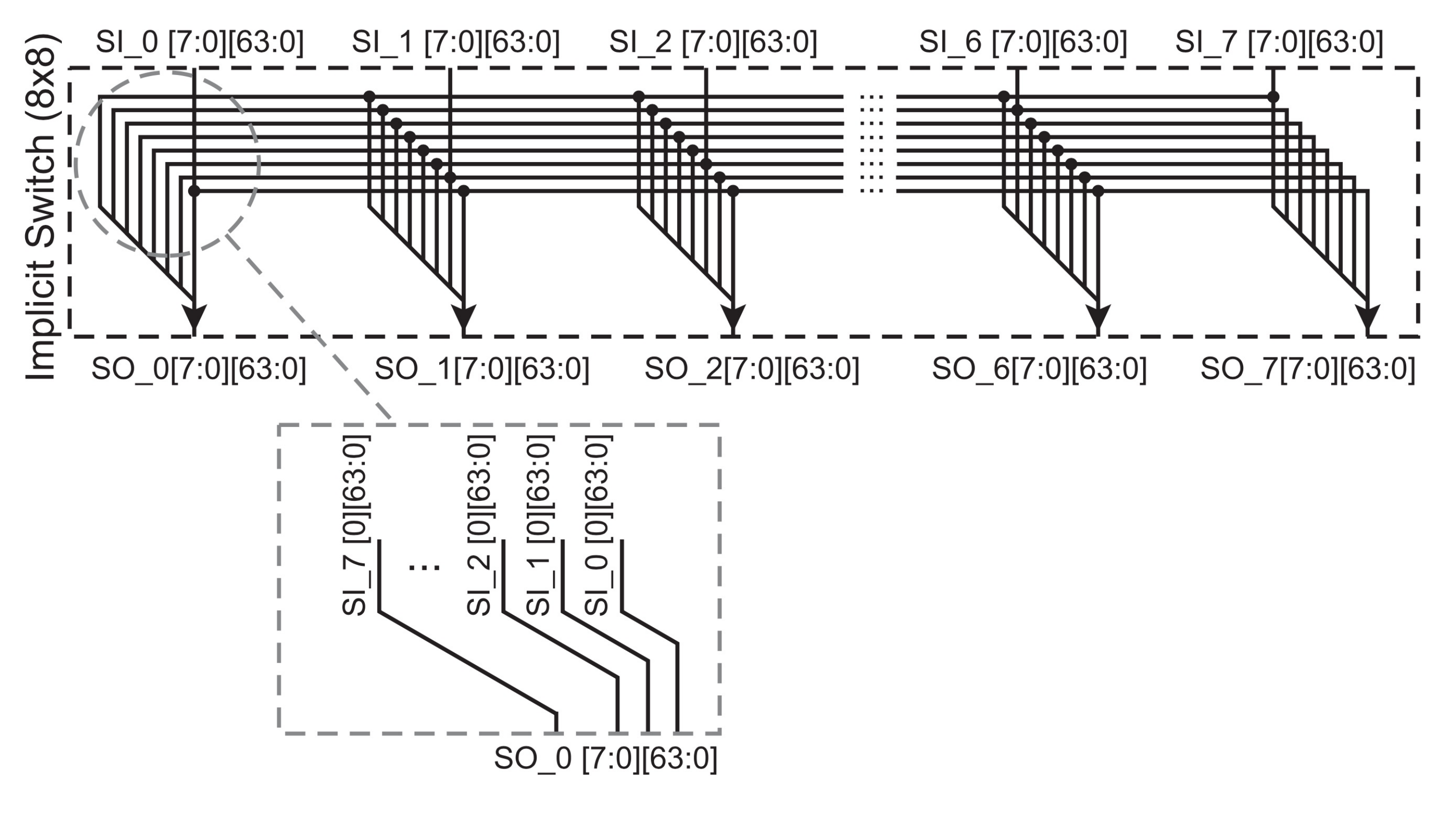}
	%\vspace{-1.25em}
	\caption{Implicit switch, 8x8~\cite{norollah2019rths}.\Amir{Faeze, please redraw this figure and share the file with us}}
	\label{fig7}
	%\vspace{-1em}
\end{figure}

}

\begin{figure}
    %\vspace{-1.7em}
	\centering
	\includegraphics[width=0.9\linewidth]{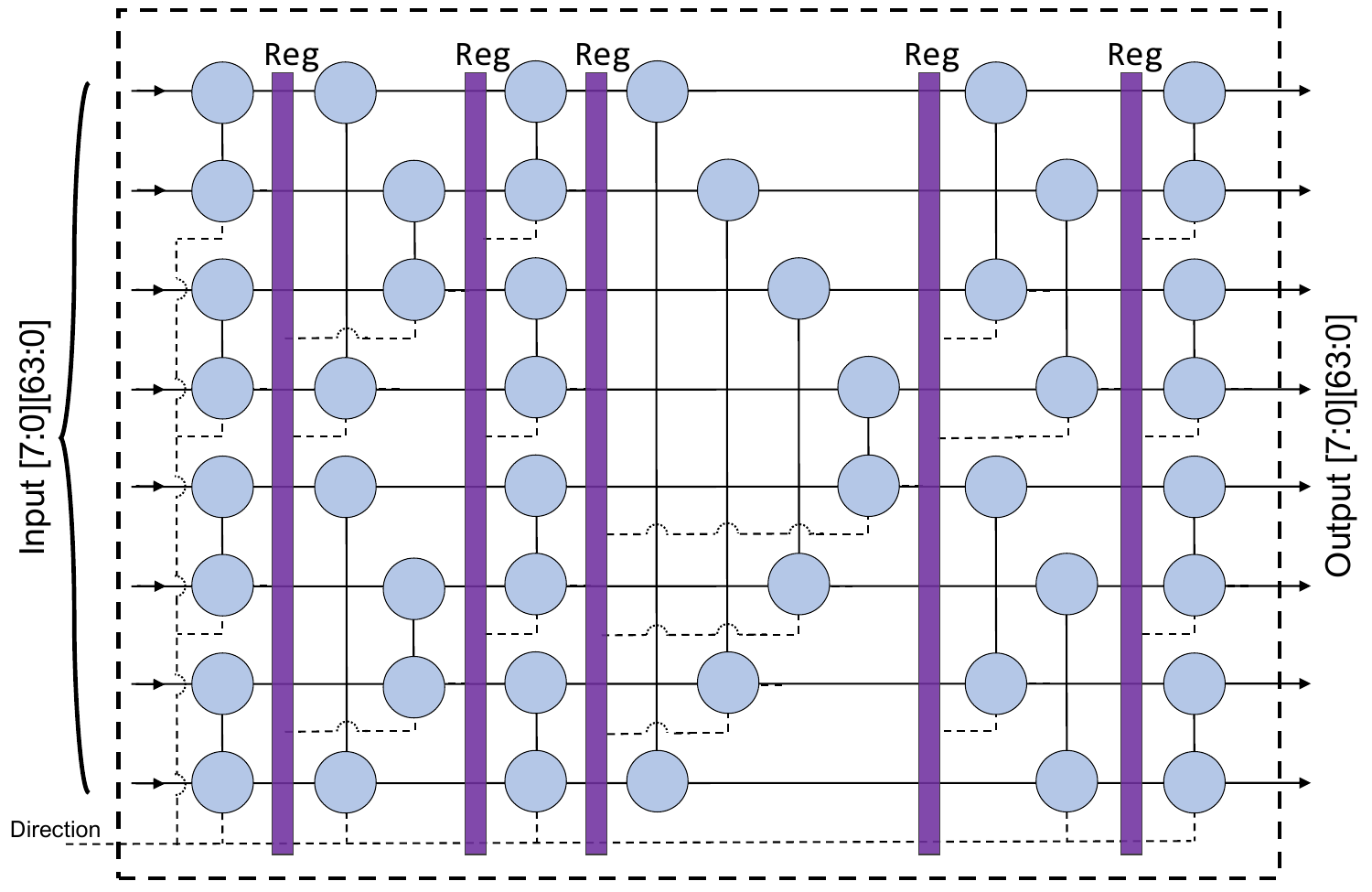}
	%\vspace{-.5em}
	\caption{Dual-mode pipeline bitonic network (DPBN) unit for 8 inputs~\cite{norollah2019rths}. The direction signal indicates the mode for sorting: normal or reverse.}
	\label{fig6}
	%\vspace{-1em}
\end{figure}

\begin{figure}
    %\vspace{-1.7em}
	\centering
	\includegraphics[width=0.70\linewidth]{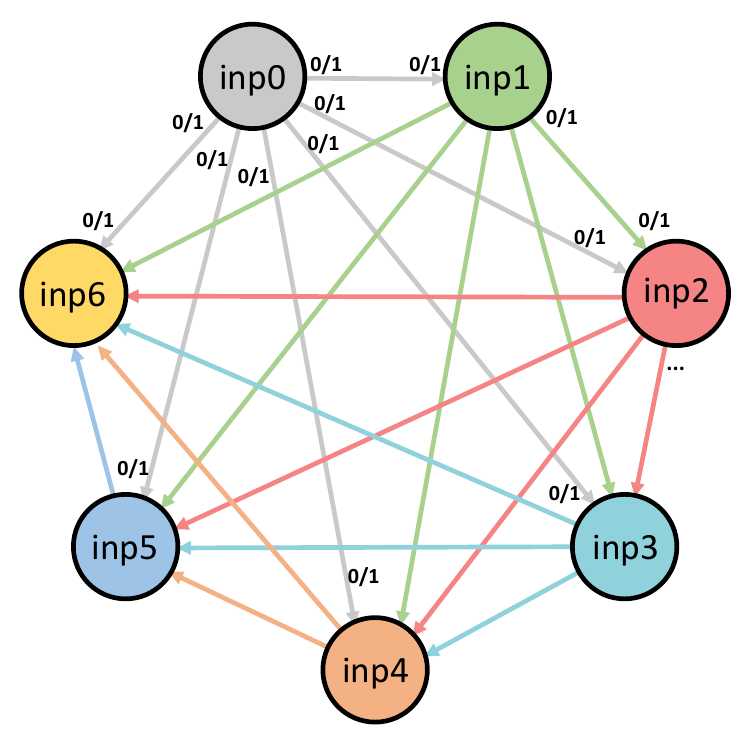}
%\vspace{-1.em}
\caption{The graph representation of~\cite{JELODARI2020103203}. Each input, represented %, symbolized 
by a vertex, is linked to all other vertices through directed edges, indicating a directed, fully connected graph. %that the graph is fully connected in a directed manner.
}
%\Amir{Can we have a better caption for this figure?}
%\Faeze{done}
	\label{fig9}
	%\vspace{-1em}
\end{figure}

%In 2020, 
Jelodari \textit{et al.}~\cite{JELODARI2020103203} proposed %presented 
a low-complexity sorting network design, which  %Their method %algorithm 
maps the unsorted input data to a graph. In this graph, %in which 
the vertices represent %are representatives of 
inputs and are fully connected through directed edges as shown in Fig.~\ref{fig9}. This structure allows comparing all inputs with each other %to one another %This results in all inputs being compared to one another 
through the directed edges connecting their corresponding vertices. At each end of any graph edge, the corresponding vertex is tagged by 0 or 1. The tags of the vertices connected by an edge are always complementary. The outgoing tag ``1''  means the source vertex is greater than or equal to the sink vertex. The sum of
the tags assigned to each vertex indicates the position of the
corresponding input data in the sorted output.

\begin{figure*}[t]
\centering 
\includegraphics[trim={0.0cm 3.5cm 3cm 1cm},clip,height=0.17\textheight]{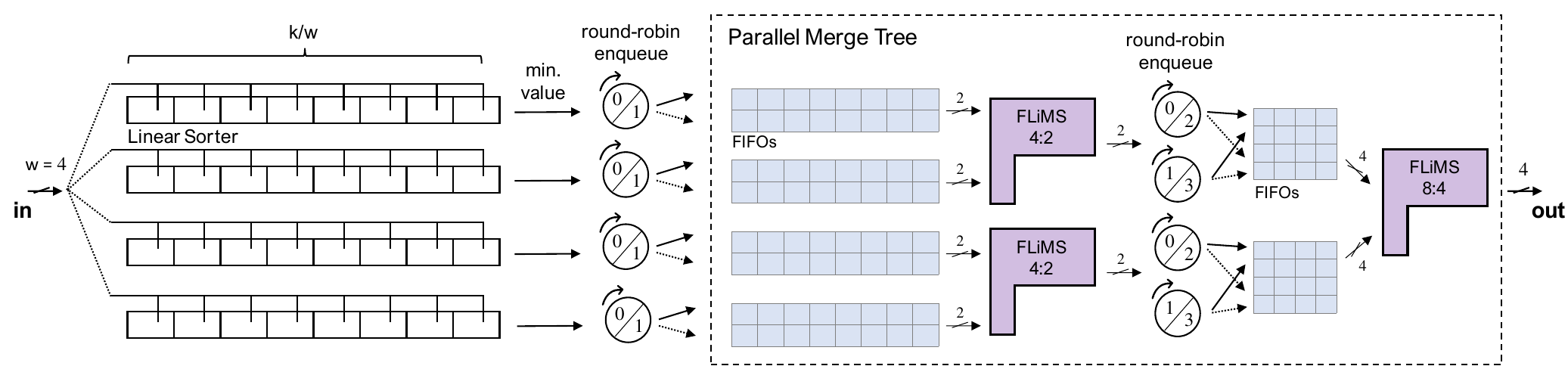}
	%\vspace{-1em}
\caption{{The high-throughput sorting system of~\cite{Papaphilippou2020AN} %not only 
 sorts data quickly while merging them %but can also merge data 
 efficiently at a rate of 4. %Additionally, it highlights that 
 The system uses a specific design where wire widths are chosen based on multiples of the width of the data values. %This design can have implications for the system's overall performance and efficiency in handling data. 
 %Initially, %it 
 %the system employs a set of ``$w$" linear sorters, where ``$w$" represents the degree of parallelism. Each individual linear sorter operates on segments of size "k/w," where "k" signifies the total merge capacity or the size of the sorted chunks. 
 This structure possesses the capability to perform both the sorting of an input containing ``$k$" elements and the merging of ``$k$" sorted lists of variable lengths. In sorting mode, a 2-bit counter value is appended to the most significant bits of all outputs from the linear sorters. This 2-bit counter is incremented whenever a new sorted chunk is flushed to the ``Parallel Merge Tree." This plays a vital role in the FLiMS (fast lightweight merge sorter) system~\cite{papaphilippou2018flims}, ensuring correct sorting prioritization for independently sorted chunks.
  }}
	\label{fig15}
% \Amir{I suggest to have a long caption similar to Fig.15 \lipsum[1]}
% \Faeze{done}
%\vspace{-1em}
\end{figure*}

\ignore{
\Amir{
Faeze,  you can start adding the remaining references. I added the bibliography of these references}
\\}

Papaphilippou \textit{et al.}~\cite{Papaphilippou2020AN},~\cite{papaphilippou2018flims} introduced %the first open-source design of 
a merge sorter tailored for small \textit{list}s, with the capability to merge sublists recursively. This feature sets their solution apart from most large-scale sorters, often reliant on pre-sorted sublists or established hardware sorter modules. Their design %aims to 
bridges the gap between high-throughput and many-leaf sorters by %employing a Verilog generator script to create 
merge sorters, allowing customization of bandwidth, data, and payload width. They assess the applicability of their solution in their specialized in-house context, specifically for database analytics. This involved %Their objective involves 
calculating the count of distinct values per key (group) from a dataset comprising key-value pairs. %To achieve this, 
They integrated a fully-pipelined high-throughput stream processor %is 
seamlessly 
%integrated 
with the sorter's output, enabling real-time result generation. 
Their streamlined process eliminates the need for temporary data storage, exemplifying task-pipelining for efficient data processing. Fig.~\ref{fig15} shows their setup. They incorporate a fast lightweight merge sorter (FLiMS) as a key component within their parallel merge/sort tree. The FLiMS unit combines two separate -already sorted- lists. The design is characterized by $w$ linear sorters, where $w$ signifies the degree of parallelism being employed. Each individual linear sorter has a length of $k$/$w$, with $k$ representing the total merge capacity or the size of the sorted chunk. This architectural arrangement %not only proficiently 
sorts an input dataset comprising ``$k$'' elements while %but also 
adeptly merging already sorted lists of varying lengths. %This versatility makes it a robust design.

%\Amir{I added the following paper's info in  TABLE II}
%\blue{
Preethi \textit{et al.}~\cite{Preethi2021L} investigated the use of
clock gating technique %a low-power technique like clock gating 
to design low-power sorters. The \texttt{bubble sort}, \texttt{bitonic sort}, and \texttt{odd-even sorting} %techniques 
algorithms are redesigned to make them low-power using the clock gating technique. The implementation results showed that clock gating can reduce the dynamic power consumption of sorters by 47.5\% with no significant impact on the performance.

\subsection{Comparison-free}
%\Amir{Good comment. I added introductory info. }
%{\color{green}{Sercan: While reading this section, I feel that I need introductory information about this sorting style, maybe a little technical, thereby transiting into the following literature analyses. If you want, I can write, or the responsible author of this section may write better than me with proper keywords and syntax}}
\label{Sec:Compfree}

%As we discussed, %previously mentioned, non-comparative 
Comparison-free sorting designs %algorithms are sorting techniques that 
do not involve direct element comparisons. Instead, they employ alternative approaches to accomplish efficient sorting. In recent years, there has been a notable surge in research and development %related to 
of this type of sorting. In this section, we will provide an overview of these advancements.

\begin{figure}
    %\vspace{-1.7em}
	\centering
	\includegraphics[trim={0.5cm 0.75cm 0.75cm 0.75cm},clip,width=0.9\linewidth]{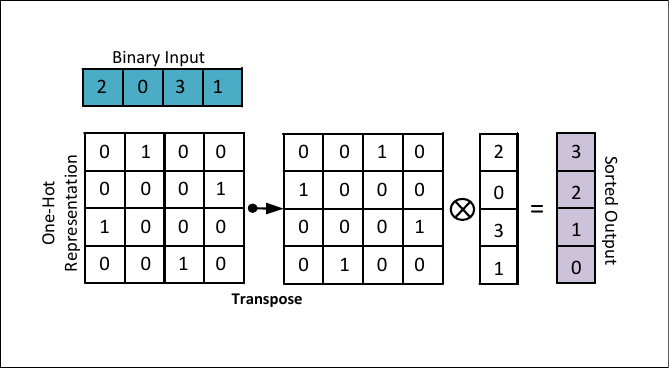}
%\vspace{-1.em}
\caption{Example of sorting four input data with the method of \cite{Abdel2017An}. }
	\label{fig2}
	%\vspace{-1em}
\end{figure}

Abdel-Hafeez and Gordon~\cite{Abdel2017An} proposed a comparison-free sorting algorithm for large datasets. %is proposed in \cite{Abdel2017An}. 
%There is The
The method operates on the elements' one-hot weight representation, a unique count weight associated with each of the $N$ elements. %no comparator in the design and 
The input elements are inserted into a binary matrix of size $N\times1$, where each element is %has a size of 
$k$ bits.  %is of size $k$-bit.
Concurrently, the input elements are converted to a one-hot weight representation and stored in a one-hot matrix of size $N \times H$. In this matrix, %where 
each stored element is of size $H$-bit and $H = N$ gives a one-hot matrix of size $N$-bit$\times$$N$-bit. The
one-hot matrix is transposed to a transpose matrix %\red{(transposed to transpose?)}
of size $N \times N$, which is multiplied by the binary matrix %—rather than using comparison operations—
to produce a sorted matrix. An example of this method is illustrated %can be found 
in Fig.~\ref{fig2}. The total number of sorting cycles is linearly proportional to the number of input data elements $N$. The architecture of \cite{Abdel2017An} is a high-performance and low-area %growth rate 
design for hardware implementation. 

Bhargav and Prabhu~\cite{bhargav2019power} later proposed an algorithm for comparison-free sorting using finite-state machines (FSMs). %the FSM module. 
Their FSM consists of six
states that describe the functionality of a comparison-free sorting algorithm dealing
with $N$ inputs. %elements.
Their proposed design shows 53\% and 68\% savings in area and power consumption compared to the design of~\cite{Abdel2017An}.

%This technique can be utilized in order to map the random output response generated by any logical design while dealing with random inputs, which can be either forced inputs by the observer or can be generated automatically. 
%\hassan{The above paragraph also needs improvement and more details. What are "write-evaluation and read-sort phases that we say they combine? Did we explain these before? What is the benefit of this method? Any improvement?}
%\Amir{Done}

%Furthermore, the design implemented for signed inputs.
%In the field of Design and Trust, it is a requirement for the designer to map the
%responses with the corresponding inputs while dealing with malicious hardware or
%information leakage. Any deviation of the responses from the required reaction must be
%recorded correctly along with its respective input. Its challenging enough to spot the
%affected output manually instead sorting the input and the responses will allow the
%designer to compare and analyses the observations.

\begin{figure}
    %\vspace{-1.7em}
	\centering
	\includegraphics[trim={1cm 0.5cm 1cm 0.5cm},clip,height=0.2\textheight]{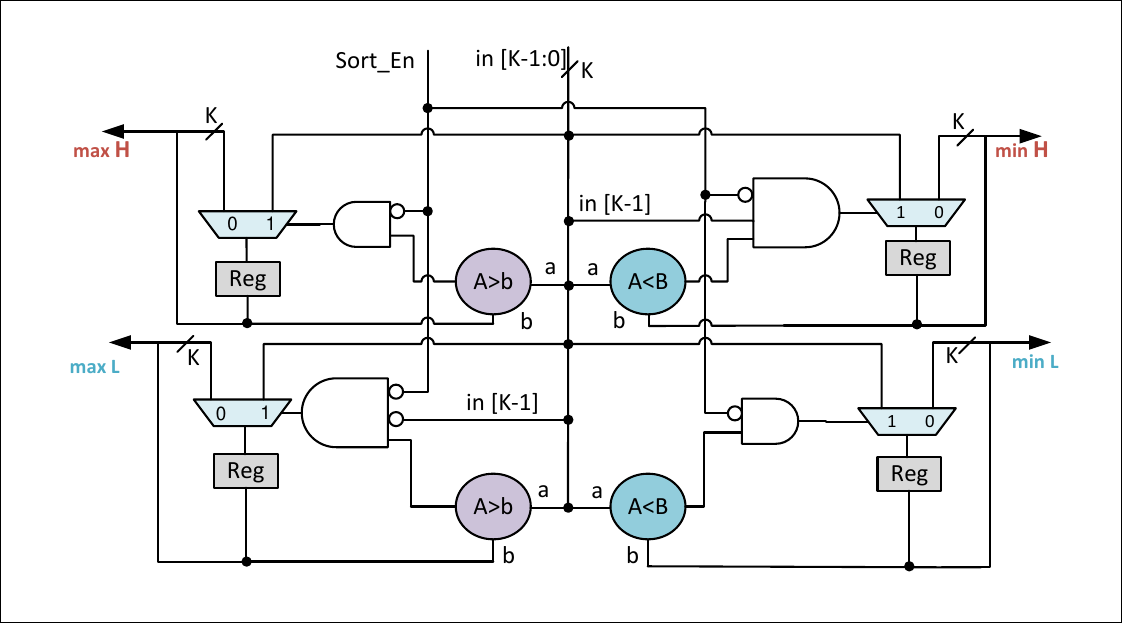}
	%\vspace{-1.em}
	\caption{Architecture of the boundary founding  module used in~\cite{Chen2021A}. 
 %\Amir{Added figure}
}
	\label{figbfm}
	%\vspace{-1em}
\end{figure}

Chen at al.~\cite{Chen2021A} improve
%\red{(improves?)} 
the number of sorting cycles, %of the previous work \red{(any specific work?)} 
%be made to further enhance performance, such as reducing
%the number of sorting cycles, 
which range from $[2N$ to $2N + 2K-1]$ to $[1.5 N$ to $2N + (\frac{2^k}{2})-2]$. %cycles. 
Their proposed architecture improves the performance of the unidirectional architecture in~\cite{Abdel2017An} by reducing the total number of sorting cycles via bidirectional sorting along with two auxiliary modules.
%\red{
One of the auxiliary modules is %the boundary finding. 
\textit{boundary finding}, %(BF)
% module 
which is designed to record the maximum and minimum values of the input data for the high-index part (max H and min H) and the low-index part (max L and min L). As %can be seen 
shown in  Fig.~\ref{figbfm},  the boundary values are stored in four $K$-bit registers where $K$ is the bit-width of input
data. %element. 
In the initial state of the circuit, the values of
max H, min H, max L, and min L are set to $2K/2$, $2K-1$, $0$,
and ($2K /2$) $-1$, respectively.
A \textit{binary finding} module shortens the range for index searching
by finding the boundaries of the range. %}
 %\red{(: Method1 and Method2)}. 
Bidirectional sorting allows the sorting tasks to be conducted concurrently in the high- and low-index parts of the 
architecture. Sri \textit{et al.}~\cite{Sri2022An} %Later in ~\cite{Sri2022An}, the authors 
reduce the area, delay, and power consumption of the design of~\cite{Chen2021A} by improving the boundary finding module.
%\hassan{What is "boundary-finding module''?}
The improvements are achieved by removing the \texttt{AND} gates and  \texttt{MUX} components.
%AND gates and multiplexers are eradicated without affecting
%the basic functionality of the module.

%, where K is the bit width of a data element.
%The architecture

Ray and Ghosh~\cite{Ray2022K-Degree} 
developed an architecture for parallel comparison-free sorting based on a model presented earlier in~\cite{ghosh2019comparison}. %.
%The proposed sorter
%and uses the base model presented in the earlier work in \cite{ghosh2019comparison}.
This work sorts $N$ data elements completely by utilizing $N$ iterations with speed-up 
of $\frac{n}{\ceil{\frac{n}{k}}  + k }$ %{\color{green}Sercan: the symbol for `rounding' seems interesting; maybe it is supposed to be both-sided and turned into to the inner side of denominator}
compared 
to non-parallel architectures.

\begin{figure}
	\centering
%trim={left bottom right top}
	\includegraphics[trim={0.85cm 1.0cm 0.8cm 0.9cm },clip,width=0.60\linewidth, %height=0.28\textwidth
	]{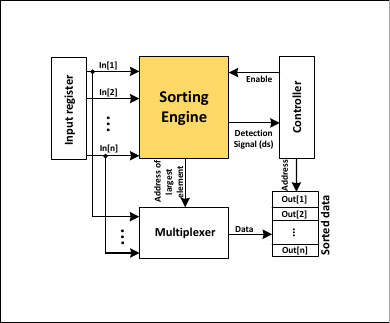}
	%\vspace{-0.5em}
	\caption{High-level architecture of the comparison-free unary sorter in~\cite{Amir2022A}.}
	\label{GeneralArchitecture}
%\vspace{-1em}
\end{figure}

\begin{figure}
%\vspace{-1.5em}
	\centering
%trim={left bottom right top}	
	\includegraphics[trim={1.3cm 1.1cm 1.1cm 1cm },clip,width=0.80\linewidth]{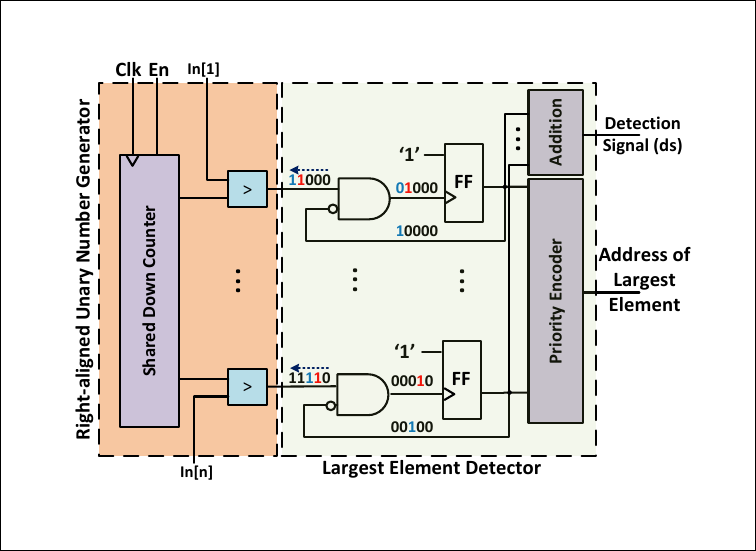}
	%\vspace{-1em}
	\caption{{The \textit{Unary Sorting Engine} proposed in~\cite{Amir2022A}. %Proposed unary processing-based sorting engine.
	}}
	\label{LED}
%\vspace{-1em}	
\end{figure}

% \Amir{DAC2022}
%Recently, 
Jalilvand \textit{et al.}~\cite{Amir2022A} proposed a fast and low-cost comparison-free sorting architecture based on UC. %unary computing. 
Similar to~\cite{jalilvand2020fuzzy, najafi2021method}, their %The proposed 
method iteratively finds the index of the maximum value by converting data to left-aligned unary bit-streams and finding the first ``1'' in the generated bit-streams.  %of the proposed comparison-free sorting unit 
Fig.~\ref{GeneralArchitecture} shows the high-level architecture. The architecture includes a sorting engine, a controller, and a multiplexer. The design reads unsorted data from the input registers and performs sorting %the sorting operations 
by finding the address of the maximum number at each step. Fig.~\ref{LED} shows the architecture of the sorting engine.
The sorting engine contains simple logic and  
%The sorting engine
converts data to right-alighted unary bit-streams. It returns the index of the bit-stream corresponding to the maximum value. This is done by finding the bit-stream that produces the first ``1''. The design also employs a controller that gets a duplication sign signal from the sorting engine and puts the next value to the output sorted register. 

Finally, Yoon~\cite{yoon2022novel} proposed a sorting engine based on the radix-2 sorting %the radix sort 
algorithm. %which 
Their sorting engine avoids comparison by creating and distributing data into buckets according to the radix-2 sorting. %Their proposed sorting engine implements the radix-2 sorting algorithm.

%\section{Rank }
\section{Partial Sorting %Methods
}

\ignore{
\Amir{Sercan would you please review the following paper? I think it is on your part.}
\red{Computer Generation of High Throughput and Memory Efficient Sorting Designs on FPGA~\cite{Chen2017Computer}
%https://ieeexplore.ieee.org/document/7930437#citations
%}
}
}

\label{Sec:PS}
Partial sorting %is the other relaxed approach to 
is primarily used to sort the top-$k$ largest or smallest values out of $N$ elements, where $k<N$. %In the literature, in the sense of surveys and research papers, very few efforts have been made. %This survey presents an explicit picture of the partial sorting technique considering the hardware-related approach. 
%As shown in Fig.~\ref{Partial}, 
Partial sortings have been used for 
%Prior works also discussed other variations of %In the literature, there are 
%partial sorting %concepts 
%such as 
determining the minimum and maximum values, finding %the decision of 
more than one relative maximum and minimum (max-set min-set selection), merging of partially sorted data, and approximate partial sorting%in approximate computing
~\cite{partialSortInitialPaper, WuTop10, GunhanSorting, Farmahini2013Modular}.
%Min max
%Selecting 
Finding the minimum and maximum values among a set of data has been particularly an important target of partial sorting. %problem that has been explored the most 
%in the literature. 
FPGA %, in particular, 
has been a popular platform for implementing this type of %partial 
sorting in hardware. %Especially FPGA-based designs have been the implementation platform that is frequently used for the realization of digital designs. 
%With a recent work in the literature, 

%\Amir{Sercan: I think it would be much better if similar to previous sections, we draw a main architecture in some of the articles. You can ask Faeze to help you with this.}
%\sercan{Sure, Amir. I drew and added one (Fig 14). Maybe I will add one more for pure partial sorting example.}

%TEMPORARY
%\clearpage
%\newpage

\ignore{
\begin{figure}
	\centering
%trim={left bottom right top}
	\includegraphics[trim={8cm 4.5cm 8cm 4.5cm },clip, width=0.95\columnwidth]{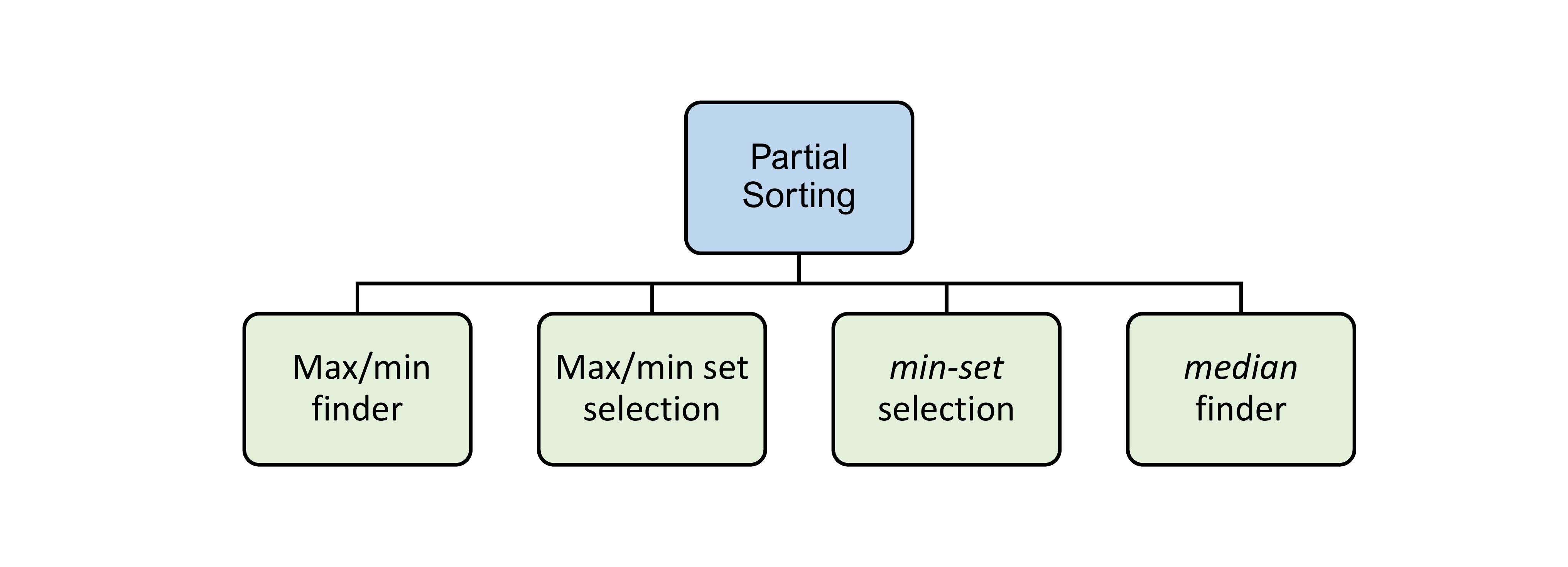}
	%\vspace{-1em}
	\caption{The goals of partial sorting.
 %\Amir{Redesigned}
 }
	\label{Partial}
%\vspace{-1em}
\end{figure}

}

\begin{figure*}[t]
	\centering
	\includegraphics[trim={0cm 7.7cm 13cm 0.5cm},clip,width=0.71\textwidth]{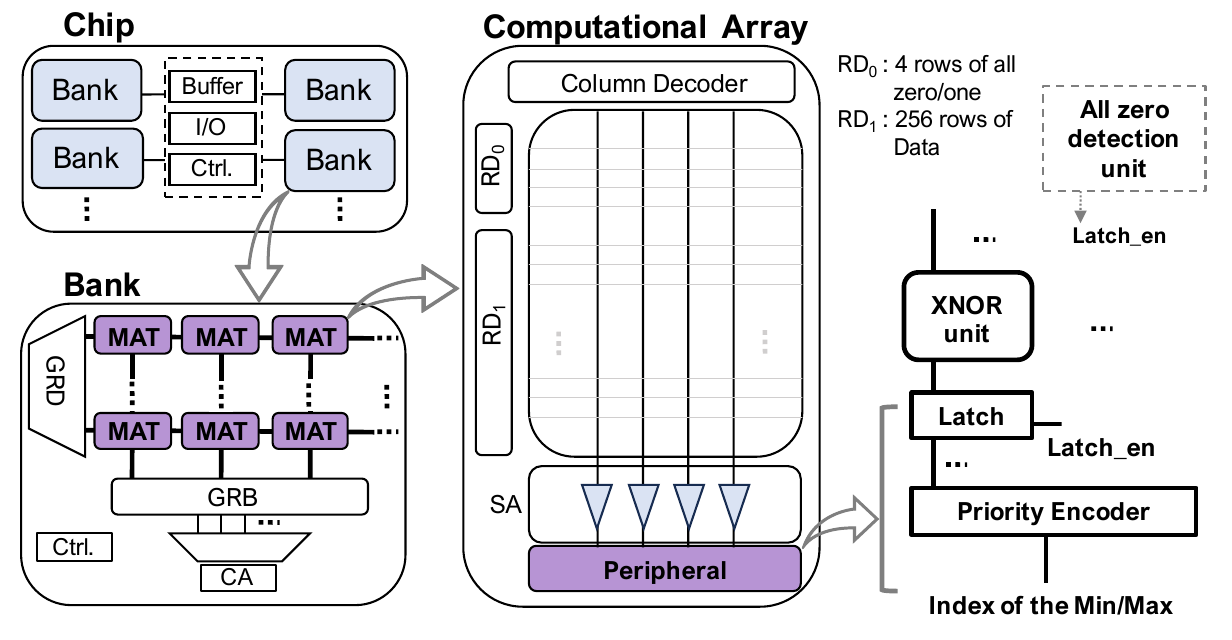}
\vspace{-1.em}
\caption{The MIN-MAX PIM architecture proposed by Zhang \textit{et al.}~\cite{AngiziPIMmax} preserves the original memory hierarchy with each DRAM chip divided into multiple banks. These partitions share the Input-Output and buffers. Each bank comprises multiple %normal 
memory matrices (MATs), which are essentially DRAM subarrays. %The authors emphasize that their 
The design supports Ambit~\cite{8686556, angizi2019accelerating} logic and instructions with enhanced support for the Dual Row Activation (DRA) mechanism, thus providing compatibility with \texttt{XNOR} operations. %Additionally, 
The Computational Array includes (i) two-row decoders, (ii)  one column decoder, (iii) modified logic sense-amplifier, (iv) one latch per bit-line, (v) pseudo-\texttt{OR} gate, and (vi) one priority encoder. The circuit for zero detection %in the unit 
uses a pseudo \texttt{OR} gate. This gate is employed to govern the update of the matching vector latch. The priority encoder returns the index of the minimum or maximum value for partial sorting purposes.}
\label{fig16}
 %\Amir{also we need  a rather long caption here :)}
 %\vspace{-1em}
\end{figure*}

Yan \textit{et al.}~\cite{minMaxFPGA} proposed an architecture for determining the $k$ largest or smallest numbers on FPGA. %by partial sorting 
%in the FPGA environment. 
%That work Yan \textit{et al.} 
Their work allows selecting %the selection of 
two min/max subsets with a real-time hardware partial sorter (RTHPS) structure 
%\hassan{Is RTHPS proposed in Yan \textit{et al.} work?}
%\sercan{yes, dr. Najafi, I combine the sentences in a proper way to make it more clear}
%\hassan{Thanks, Sercan.}
consisting of even-odd swap blocks, a bitonic sorting network, and parallel swap blocks. 
%Speaking of even and odd numbers, vector-based odd-even merge sorting by Korat \textit{et al.} is also worth mentioning \cite{8249010}. 
Korat \textit{et al.}~\cite{8249010} proposed a sorting algorithm that partially sorts the odd and even parts in a vector structure. Their method guarantees a linear time complexity with $O(n)$. %is guaranteed. 
The hardware unit includes two multiplexers and a comparator, which is responsible for ordering input pairs. Their FPGA-based hardware design implemented on a Xilinx VIRTEX-7 VC707 FPGA consumes 136 LUTs and 181 registers with a working frequency of 370 MHz when sorting eight inputs.

Median sorting is another practice of partial sorting with wide application %. The median calculation is an important process 
in image processing, particularly for image enhancement. Various hardware designs for median filtering have been proposed in the literature. Subramaniam \textit{et al.}~\cite{SubramaniamMedian} proposed a hardware design for finding the median value of a set of data. They employ selective comparators as a means to locate the median, allowing for partial sorting with fewer elements compared to the conventional designs that necessitate a fully sorted list.
%They utilize selective comparators, a unit for finding the median. Thus, partial sorting is performed with fewer elements compared to the architecture requiring a fully sorted list in conventional designs.} 
%\hassan{What "selective" means here? fewer elements compared to what?}
%\sercan{I have gone over it, Dr. Najafi.}
%Thanks
CAS operations are obtained using a comparator and two 2-to-1 \texttt{MUX}s. They implement the design on an FPGA (Xilinx FPGA Virtex 4 XC4VSX25) %device 
and evaluate it using an image processing case study~\cite{SubramaniamMedian}. %The median calculation is an important process in image processing applications, and its hardware-designed block is implemented in the scope of partial sorting. 
%Median filtering occurs in many image processing where sliding window filtering operations and image enhancement processes are vital. 
%The performance is measured in image processing applications. 
Using a pipelined architecture, Cadenas \textit{et al.}~\cite{7065320, Cadenas2} proposed a median filtering architecture using accumulative parallel counters. Najafi \textit{et al.}~\cite{Najafi2018Low-Cost} further implemented a low-cost median filtering design based on UC by converting data to unary bit-streams and processing them in the unary domain using simple standard \texttt{AND} and \texttt{OR} gates. Finally, Riahi Alam \textit{et al.}~\cite{alam2022sorting}  proposed a binary and a unary %in-memory 
architecture for energy-efficient median filtering completely in memory. %using memristive devices.

%PIM min max
%The use of 
Finding the maximum and minimum values is one of the current topics in in-memory computing applications. %technologies. %within the scope of partial sorting in the literature. 
Zhang \textit{et al.}~\cite{AngiziPIMmax} proposed an in-memory min-max sorting architecture in DRAM technology for % that can be used in 
fast and big data applications (see Fig.~\ref{fig16}). Sorting and graph processing applications are provided with an architecture that %gives 
produces results 50 times faster than a GPU. %Zhang \textit{et al.} 
This architecture includes two-row decoders, a one-column decoder, a modified logic sense amplifier with a typical sense amplifier (TSA), one latch per bit-line, a pseudo-\texttt{OR} gate, and one priority encoder (for the resultant index of minimum and maximum locations).

%Digital
Campobello \textit{et al.}~\cite{CAMPOBELLO2012178} discuss sorting networks' complexity and propose a %digital design of a 
multi-input maximum finder circuit. Their design finds the maximum value by using an \texttt{XNOR} comparator, a zero catcher (via $Q$-port feedbacked D flip-flip), a buffer with enable for each input, an \texttt{OR} gate, and a D-flip-flop. %hardware at the output gives the maximum value. 
%Application
%When we look at the application-based sorting algorithms and the partial sorting concept, we see that the most important circuit design is the maximum and minimum finder designs. For example, 

\begin{figure*}[t]
\centering 
\includegraphics[trim={7.5cm 8.8cm 7.3cm 1.5cm},clip,height=0.23\textheight]{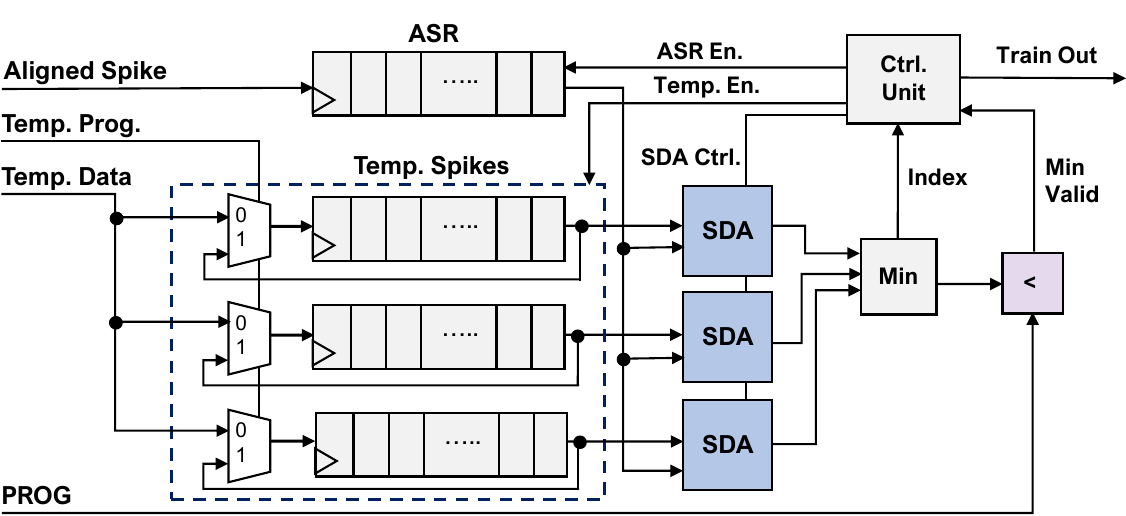}
%\vspace{-0.5em}
	\caption{{Template matching-based architecture for spike sorting. The aligned spike is directed into an ASR (Aligned Shift Register) module, which has been set up for parallel input and serial output. The values stored in the templates and the ASR module are then transferred into some SDA (Squared Difference Accumulator) units. These SDA units are used to calculate and accumulate the squared differences between the spike waveform preserved in the ASR and templates. The MIN unit identifies and conveys the minimum value to the comparator, along with the index of the minimum value, which is then passed on to the Control Unit. The substantial reduction of raw data to sorted spikes is achieved by transmitting only those sorted spikes (in a partial sorting manner) that match a small set of frequently encountered waveforms~\cite{spikingMIN, 8675427}. 
 %\Amir{Redesigned}
 }}
	\label{spike_sort}
%\vspace{-1em}
\end{figure*}

 Partial sorting can also be used as an intermediary tool to help understand data, \textit{e.g.,} to find outliers~\cite{Mitra2008}. This includes the complex task of \textit{spike sorting} in brain-inspired computing.
 %Sorting %term 
%appears in brain-inspired computing  as a more complex task in %the form of %named \textit{i.e.}, 
%\textit{spike sorting}. %The spike processing applications require sorting. 
%the sorting operation, 
%which could be a potential avenue for new %the new-coming 
%research.  
Spike sorting encompasses algorithms designed to identify individual spikes from extracellular neural recordings and classify them based on their shapes, attributing these detected spikes to their respective originating neurons. This sorting process differs %somewhat
from conventional sorting as it %spike sorting 
involves machine learning-related steps such as detection, feature extraction, and classification. %Therefore, 
Instead of straightforward scalar sorting, spike sorting resembles the segmentation of patterns within brain signal pulses \cite{spikeMWSCAS}. Spike sorting involves partial sorting for tasks such as %From a hardware-assisted sorting perspective, spike sorting presents a compelling application, as it may involve partial sorting for tasks like 
early learning termination, outlier analysis, and spike activity thresholding. In the literature, spike sorting for a unit activity may %also
encompass partial sorting to separate multi-unit activity into distinct groups of single-unit activity~\cite{6105476}. %Hence, in addition to the primary objectives of partial sorting outlined in Fig.~\ref{Partial}, 
The segmentation of spike data plays a significant role in distinguishing specific activities within the overall spike data. Within the realm of spike processing, some studies underscore partial sorting for outlier analysis of the spikes \cite{Mitra2008}, while others commend it for thresholding operations \cite{Christie2014}. %The anticipated sorting hardware development is forecasted to be in brain-inspired computing. 
 Valencia and Alimohammad~\cite{spikingMIN, 8675427} implement a %need a minimum
hardware module for spike sorting architecture. {Their design incorporates a template matching unit to compute the minimum distance between spikes during the spike sorting process. Fig.~\ref{spike_sort} depicts the spike sorting design, which relies on template checks and minimum distance calculations.
%Regarding future prospects, the anticipated 
Such advancement in hardware-powered sorting %is expected to 
%aligns well and 
is expected to open new research avenues in emerging machine-learning models, particularly brain-inspired computing. %with future research on brain-inspired computing and . 
%Spike processing, which necessitates partial sorting for rapid and efficient data processing like early decision termination, holds promising potential as a focal point for emerging machine learning research in the literature.
}

%based on template matching. 
%Yu \textit{et al.}~\cite{6132381} propose a low-cost hardware architecture for spike-sorting.
%Another spike-sorting study is offered by Yu \textit{et al.}~\cite{6132381}. 
%The Hebbian eigenfilter algorithm eliminates covariance analysis and eigenvalue decomposition for principle component analysis (PCA). %Thus, a low-cost hardware architecture is proposed. 
%In-brain spike sorting is another %the most 
%common application. 
%Liu \textit{et al.}~\cite{7761590} implement a hardware design for another common application, %the general framework of the 
%in-brain neural spike sorting. 
%\hassan{Is the design of Liu et al a partial sorting design?}

%\sercan{Dr. Najafi, spike sorting in a general sense, is explained in some works as a segmentation of single unit task of multi-unit task in a pulse train so I generally added this application under the partial sorting. A further explanation is well given in this work: \href{https://ieeexplore-ieee-org.ezproxyprod.ucs.louisiana.edu/stamp/stamp.jsp?tp=&arnumber=6105476}{LINK}, Page:2, right column, end of the first paragraph: " In such cases, spike sorting, the process of separating multi-unit activity into groups of single unit activity, is necessary". Also, the partial spike sorting term appears in some publications for thresholding and spike outliers. Spike sorting is a little different than the scalar sorting. This term is used for the pattern segmentation in long pulse trains as spikes. This includes feature extraction and classification in hardware. }
%\hassan{Thanks Sercan for adding this interesting application of partial sorting.}

%---------------------------------------

\section{In-Memory Sorting}
\label{Sec:InMemSorting}
{
In traditional processors, %processing, 
data are retrieved from disk storage and loaded into memory for processing.
In this conventional 
%design 
approach, % Von-Neumann approach, 
a significant portion of the total processing time and energy consumption is wasted for transferring data between the memory and processing unit. 
}
%The 
Most prior %software-based 
sorting designs are implemented based on this Von-Neumann architecture with separate memory and processing units~\cite{194088}. 
In-memory computation (IMC) --aka processing-in-memory (PIM)-- is a promising solution %paradigm for solving 
to address this data movement %Von Neumann
bottleneck. In this processing approach, %paradigm,
the chip memory is used %responsible %the memory chips of the hardware have the capacity 
for both storage and computation~\cite{trancoso2015moving}. %will add more citations later.
%Similarly, 
To address the data movement issue and improve sorting speed, %issue of the %decrease the complexity of the 
%traditional sorting
%sorting 
%designs, 
\textit{in-memory sorting}~\cite{alam2022sorting, prasad2021memristive, yu2022fast, 9771078} has been proposed.
In particular, the special properties of %the Memristive storage as a 
 non-volatile memories (NVMs) make
them %it 
a promising candidate for efficient sorting in memory. % the efficient execution of sorting %different 
%tasks 
%(\textit{e.g.}, sorting) 
%in memory. %with high storage density and IMC capability.

Chu \textit{et al.}~\cite{chu2021nvmsorting} proposed an NVM-friendly sorting algorithm called ``NVMSorting". NVMSorting %that 
is a modification of the MONTRES algorithm~\cite{7932112}, %(\red{maybe a 1-2 sentence clue on the algorithm}) 
%\blue{Newsha:
%MONTRES is a sorting 
{%Newsha: 
a sorting algorithm resembling merge sort, designed for flash memory.
MONTRES aims to enhance performance by minimizing I/O operations and reducing the generation of temporary data during sorting. It includes a run generation phase and a run merge phase, employing optimized block selection, continuous run expansion, and on-the-fly merging for efficient data organization.}
%~\cite{7932112}.
%} and 
NVMSorting has the ability to detect partially ordered runs by using a new concept, called \textit{natural run,}
%\hassan{What "run" means here?}
to reduce the sorting cost. A natural run consists of multiple blocks. The items within each block are not required to be sorted, but the items between any two consecutive blocks are ordered. In the first step, the algorithm searches for the partially ordered runs (\textit{i.e.,} natural runs) in the input data. %which are %also . 
The next step is the
\textit{run generation}, which is based on %the 
a merge-on-the-fly mechanism and a run expansion mechanism. DRAM is divided into two sections: I) workspace for the natural runs, and II) workspace for the other input data. Chu \textit{et al.} %They 
take advantage of the NVM's byte-addressable capability to merge the runs. Their evaluations show that %They demonstrated that 
NVMSorting is more efficient than the traditional merge sorting algorithms in terms of execution time~(t) and number of NVM writes~(w).
%performance and latency. %\blue{The performance metrics encompass the execution time (t) and the number of non-volatile memory (NVM) writes w.}
%\hassan{what "performance" means here? cannot we just say latency?}
However, if the dataset is %entirely randomly 
entirely random, %and unsorted, 
NVMSorting can achieve similar performance to MONTRES, hybrid sort, and external sort~\cite{viglas2014write}.
%\hassan{Is [56] a reference for hybrid sort or external sort or both? }
%\hassan{Any reference for hybrid sort and external sort? }
%\Newsha{DONE.}
%\hassan{They did not provide a reference for hybrid sort and external sort in their result section because they probably provided those references somewhere else in the paper. Do they provide any reference the first time they mention hybrid sort and external sort in the paper? The reader of our paper has no idea what you mean by hybrid sort and external sort.}

\begin{figure}
    %\vspace{-1.7em}
	\centering
	\includegraphics[trim={0.5cm 0.2cm 0.5cm 1.3cm},clip,width=0.8\linewidth]{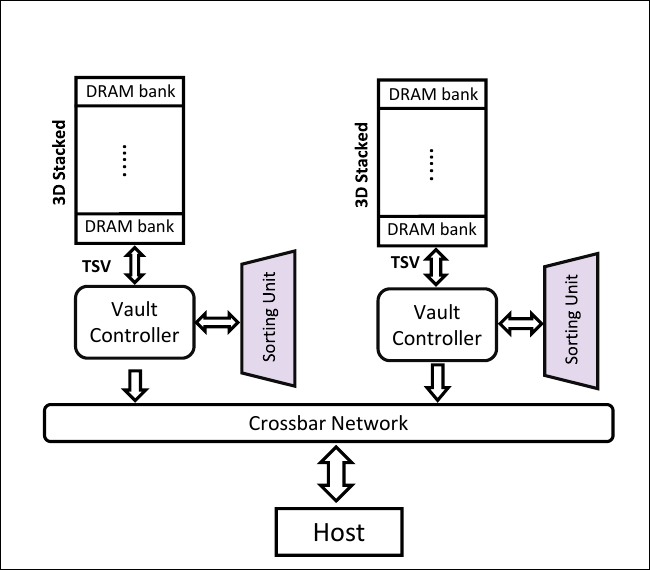}
	%\vspace{-1.em}
	\caption{Overall architecture of the IMC-Sort. A single stack HMC vault is composed of several DRAM banks that are linked to the logic layer via through-silicon vias (TSV)~\cite{li2020imc}.}
	\label{fig13}
 %\Amir{Please cite the reference in the caption of the figure}
	%\vspace{-1.5em}
 %\hassan{This figure is not cress-referenced in the main text.} \Newsha{Newsha: added the reference in the text.}
% \hassan{Where?}
\end{figure}

Li \textit{et al.}~\cite{li2020imc} %, the authors 
proposed a PIM architecture called IMC-Sort to perform parallel sort operations using a %the 
hybrid memory cube (HMC). As shown in Fig.~\ref{fig13},
IMC-Sort is comprised of sorting units that are specifically designed to operate within each HMC vault's logic layer. %Additionally, 
The control unit of the HMC vault is %has been
enhanced with %the required 
some logic to carry out the sorting process. The sorting units in IMC-Sort are capable of parallel access and utilize the HMC crossbar network to communicate with one another. 
In %the 
an ``\textit{Intra-vault merging}" step, %approach,
they utilize a \textit{chunking} technique to accommodate a range of input sequence lengths using a fixed number of CAS units and a fixed input permutation unit. They divide %that divides 
the sequence into %some 
chunks of a specific size determined by the number of CAS units. Then, they sort the chunks. %and then sorting the chunks.
Finally, the sorted values are merged into a single sorted sequence. On the other hand, an ``\textit{Inter-vault merging}" step %approach 
combines the sorted values or sequences from all vaults to produce a globally sorted sequence. IMC-Sort delivers %a 
16.8$\times$ and 1.1$\times$ speedup and %a
375.5$\times$ and 13.6$\times$ reduction in energy consumption compared to the widely used CPU implementation and a state-of-the-art near memory custom sort accelerator, respectively~\cite{farmahini2012modular}, ~\cite{pugsley2015fixed}, ~\cite{samardzic2020bonsai}, ~\cite{zhou2016high}, ~\cite{srivastava2015hybrid}.

%\Newsha{DONE}
%\hassan{Check the paper again to find out to what design/paper they are referencing when they say "state-of-the-art near memory custom sort accelerator". They are saying IMC-Sort is faster and consume less than a near memory custom sort accelerator. What is this accelerator? They should have cited a paper for this.}

Riahi Alam \textit{et al.}~\cite{alam2022sorting}
%In~\cite{alam2022sorting}, they 
proposed the first in-array (in-memory) architectures for high-performance and energy-efficient data sorting completely in memory using memristive devices.
They introduce two different architectures. The first architecture, ``\textit{Binary Sorting}," is based on the conventional weighted binary representation, while the second architecture, ``\textit{Unary Sorting}," is based on the non-weighted unary representation. Both of these sorting designs achieve %have 
a significant reduction in the processing time compared to prior off-memory binary and unary sorting designs. The memristor technology they used is based on the \textit{stateful logic} in which the input and output are both presented as the state of input and output memristors. In stateful logic, values are stored and maintained within memristive switches through their resistance states. These switches not only store logic values but also perform logical operations, exhibiting both memory and computational capabilities~\cite{borghetti2010memristive, kvatinsky2014magic,gupta2018felix}. 
%\blue{Newsha: ~\cite{}, , ~\cite{taherinejad2021sixor } , ~\cite{}}. 
They implement the boolean operations with memristor-aided logic (MAGIC)~\cite{kvatinsky2014magic}
%\blue{Newsha: refrence 32 at the end of the sentence} 
in a crossbar implementation. Each MAGIC logic gate utilizes memristors as inputs, which contain previously stored data, and additional memristor functions as the output. %With that being said, 
Parallel architectures such as CAS-based sorting networks can be executed efficiently %parallelly 
within the memory using these IMC logic operations~\cite{alam2022sorting}.

In the first design, the memory is split %authors split the memory 
into multiple partitions to enable parallel execution of different CAS operations of %in 
each bitonic CAS stage. The number of partitions indicates the number of CAS units that can run in parallel. The first two inputs of each partition are sorted using a basic sorting operation. Then the maximum value of each basic sort operation is copied to another partition %which is 
determined by the sorting %bitonic 
network. The second design is a complete unary sort system that follows the same approach as the binary implementation but represents and processes the data %is 
in the \textit{unary} domain with uniform unary bit-streams% of first all 1s and then all 0s~
\cite{najafi2018low}. %poppelbaum1987unary
%Therefore
The %basic 
comparison operations are %should be
implemented in this design based on %the 
a basic unary sorting unit. Their performance evaluation results show a significant latency and energy consumption reduction compared to the conventional off-memory designs. On average, their in-memory binary sorting resulted in a 14× reduction in latency and a 37× reduction in energy consumption. On the other hand, the average latency and energy reductions for the in-memory unary sorting design were much greater, at 1200× and 138×, respectively.
Further, they implemented two in-memory binary and unary designs for Median filtering based on their developed in-memory basic sorting units. Their results showed an energy reduction of 14$\times$ (binary) and 5.6$\times$ (unary) for a \linebreak 3 $\times$ 3-based image processing system, %(in-memory binary and unary designs, respectively)  
and 3.1$\times$ and 12$\times$ energy reduction for binary and unary median filtering, respectively, for a 5 $\times$ 5-based image processing system %, results showed  (in-memory binary and unary designs, respectively) improvements, 
compared to their corresponding off-memory designs.

Today's systems %, in particular, 
often face memory bandwidth constraints that can limit their performance. 
The efficiency of the sorting algorithms can be significantly impacted %influenced 
by the available % amount of  
memory bandwidth. %Nevertheless, %real-world 
To overcome the bandwidth problem in large-scale sorting applications, Prasad \textit{et al.}~\cite{prasad2021memristive} proposed an iterative in-memory min/max computation technique. %algorithm. 
They applied a novel mechanism called ``RIME'', which %as a novel approach that 
enhances bandwidth efficiency by enabling extensive in-situ bit-wise comparisons. %This mechanism 
RIME eliminates unnecessary data movement on the memory interface, resulting in improved performance. %RIME provides 
They provide an API library %further provide the users with an API library, providing them 
with significant control over essential in-situ operations like ranking, sorting, and merging. With RIME, users can efficiently manage and manipulate data. %, leading to more streamlined and optimized applications.
To perform bit-serial min/max operation, %computation, 
they execute an iterative %repetitive 
search for bit value (1 or 0) within individual columns of a data array %is done 
using %the
a 1T1R memristive %memristor 
memory. In %With 
%each search, 
each iteration of the search, 
a match vector is generated to identify which rows in the array should be eliminated from the dataset. %Therefore, 
The memory array must be capable of performing two additional operations, namely bitwise column search and selective row exclusion. 

The algorithm starts by examining the binary values of all bit positions, beginning from the most significant bit position in a set of numbers. This process is carried out using a $k$-step algorithm, during which some of the non-minimum or non-maximum values may be removed from the set at each step. At each step, a selection of matching numbers is formed by searching for ``1'' at the current bit position. The selected numbers are removed from the set only if the set and selection are unequal. This results in all the final remaining numbers in the set having the minimum value. By eliminating the unnecessary data movement for finding min/max of given data, their sorting operation obtains a bandwidth complexity of $O(N)$. 
With the suggested in-memory min/max locator, the %expenses 
costs of accessing bandwidth when searching for the $k^{th}$ value in a range of data decrease to $k$ operations, which shows a bandwidth complexity of $O(k)$. 
Their simulation results %The results of their simulations
on a group of advanced parallel sorting algorithms demonstrate a significant increase in throughput ranging from 12.4$\times$ to 50.7$\times$ %times 
when using RIME. %\blue{the explanation is added.}
%\hassan{What is RIME? Any reference?}

\begin{figure}
    %\vspace{-1.7em}
	\centering
	\includegraphics[trim={14cm 1.7cm 13.7cm 1.7cm},clip,width = 0.95\columnwidth]{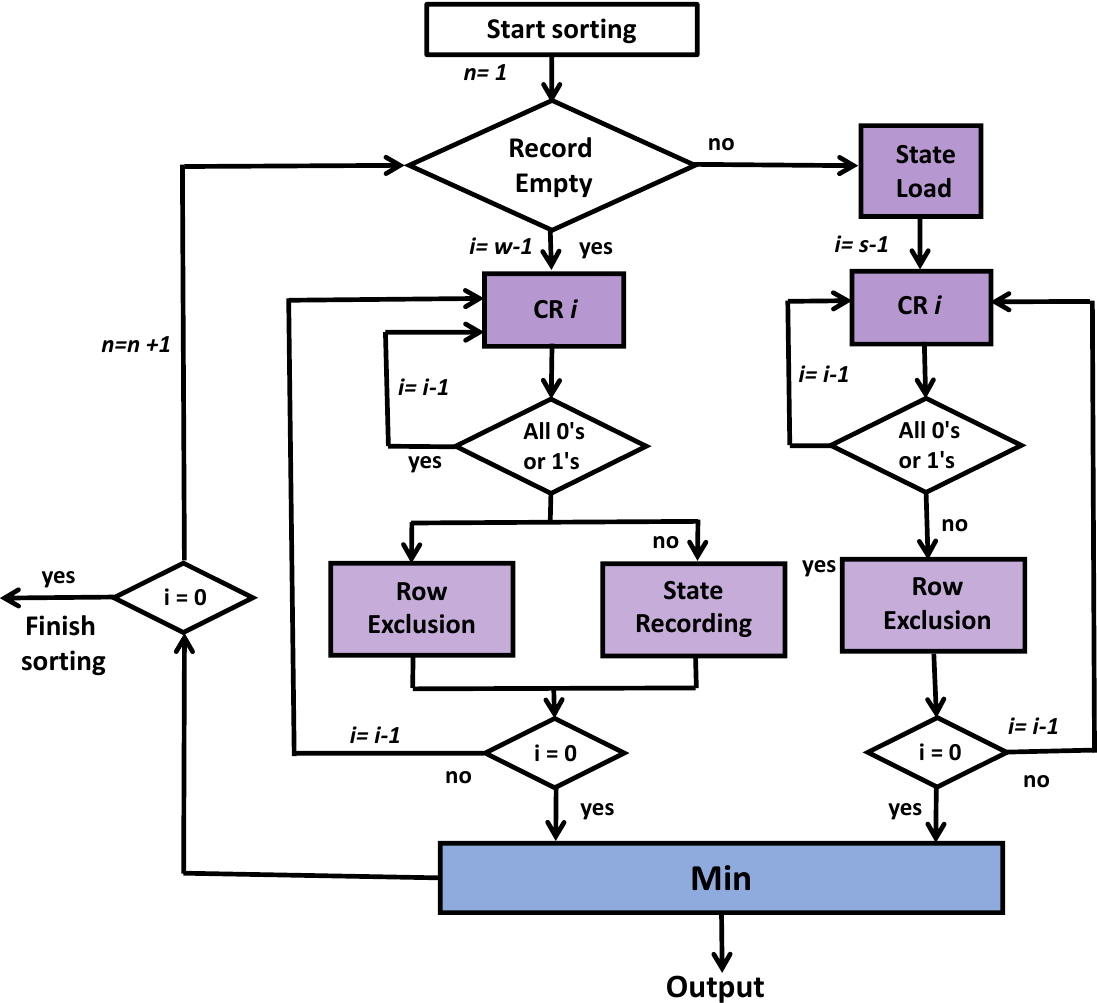}
	%\vspace{-1.em}
	\caption{Iterative min search with proposed column-skipping algorithm~\cite{yu2022fast}.}
	\label{fig14}
	\vspace{-0.5em}
\end{figure}

Yu \textit{et al.}~\cite{yu2022fast} improve the speed and performance of Parasad \textit{et al.}'s design by proposing %the last work. 
%\hassan{By "last work" do you mean Parasad \textit{et al.}'s work?}  \blue
%{yes}
%They propose 
a column-skipping algorithm that keeps track of the column read conditions and skips those that are leading 0’s or have been processed previously (see Fig.~\ref{fig14}). %The
A \textit
{bank manager} enables column-skipping for datasets stored in different banks of the memristive memory. For detecting and skipping redundant column reads the algorithm records the $k$ most recent row exclusion states and their corresponding column indexes, which can be reloaded to avoid repeating these states.

To tackle the sorting challenges of large-scale datasets, Zokaee~\textit{et~al.}~\cite{zokaee2022sky} proposed %another approach called 
Sky-Sorter, a cutting-edge sorting accelerator powered by Skyrmion Racetrack Memory (SRM). %To achieve this objective, the author 
Sky-Sorter leverages the unique capabilities of SRM, enabling the storage of 128 bits of data within a single racetrack. Sky-Sorter adopts the \texttt{sample sort} algorithm, which encompasses a sequence of four essential steps: sampling, splitting marker sorting, partitioning, and bucket sorting. 
First, it employs a random sampling technique to estimate the distribution of the dataset. This sampled subset is then sorted, and specific records are selected as splitting markers. The markers are crucial for defining the boundaries of non-overlapping buckets. The next step involves partitioning, where all records, excluding the splitting markers, are allocated to appropriate buckets based on their relationship to the markers. Lastly, each bucket is sorted individually, and the results are concatenated to produce the final sorted sequence. Bucket sorting, known for its high parallelizability, is the key to this algorithm's efficiency, with the distribution of bucket sizes playing a crucial role in maintaining balance. To achieve balanced distribution and prevent load imbalances during bucket sorting, it is essential to distribute records evenly across all buckets. Larger random sampling sizes contribute to more accurate estimates of the data distribution and less variability in bucket sizes. The algorithm ensures that the probability of any bucket exceeding an upper size limit is nearly zero. In rare cases where a bucket size surpasses this threshold, the algorithm triggers the resampling of splitting markers to maintain uniformity in bucket sizes. The fundamental cell structure of SRM is composed of four integral parts. These components encompass two injectors devoted to the creation of skyrmions, a detector designed for the precise detection of skyrmions, a nanotrack to facilitate the controlled motion of these skyrmions and peripheral circuits that support and coordinate the functionality of the entire cell. The authors claim that Sky-Sorter improves the throughput per Watt $\sim$4$\times$ over prior FPGA-, Processing Near Memory (PNM)-, and PIM-based accelerators when sorting with a high bandwidth memory DRAM, a DDR4 DRAM, and an SSD~\cite{samardzic2020bonsai, li2020imc, prasad2021memristive}.

To address the challenges of sorting vast datasets with limited memory resources, significant efforts have been dedicated to enhancing external sorting algorithms. While efforts have been made to enhance external sorting algorithms, few have considered the I/O requests and byte-addressable characteristics of NVM. Liu~\textit{et al.}~\cite{liu2023lazysort} proposed LazySort, an external sorting algorithm tailored to the NVM-DRAM hybrid storage architecture. LazySort leverages NVM's byte-addressable feature and locally ordered data to minimize write operations to NVM. It comprises two stages: run generation and merge. %, offering evident advantages over traditional external sorting algorithms when employed with NVM. 
To enhance efficiency, they introduce an optimization strategy known as RunMerge for the merge stage. %of LazySort. 
RunMerge intelligently merges non-intersecting data blocks based on the range of an index table records, reducing the total number of runs and memory usage. To validate the performance, %of LazySort, 
they %the researchers 
established a real NVM-DRAM experimental platform and conducted comprehensive experiments. The results showed %were compelling, demonstrating 
LazySort's superior time performance and significantly reduced NVM write operations. Compared to traditional external sorting algorithms, LazySort reduced sorting time by 93.08\% and minimized NVM write operations by 49.50\%. This design then addresses an important need for efficient external sorting methods for %in the context of 
NVM-DRAM hybrid storage. %, making it a noteworthy contribution to the field.

\ignore{
\begin{figure*}
	\centering
	%\vspace{-1.em}
%trim={left bottom right top}
	\includegraphics[trim={1.2cm 2cm 3cm 13cm},clip, height=0.113\textheight]{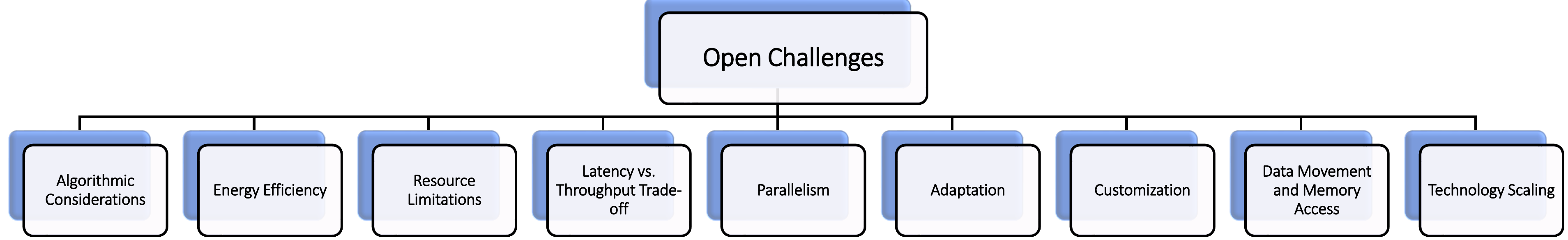}
	%\vspace{-1.5em}
 \centering
	\caption{Open challenges in hardware-assisted sorting. }
	\label{challenges}
%\vspace{-1em}
\end{figure*}

}

\begin{table*}
\centering
\caption{Various Sorting Network Configurations and their Principal Characteristics by Zuluaga \textit{et al.}~\cite{10.1145/2854150}}
%\vspace{-0.5em}
\begin{tabular}{|p{9cm}|c|c|c|c|c|} 
\hline
\textbf{Architecture} & \textbf{Logic} & \textbf{Storage} & \begin{tabular}[c]{@{}c@{}}\textbf{Storage} \\ \textbf{Type}\end{tabular} & \begin{tabular}[c]{@{}c@{}}\textbf{Streaming} \\ \textbf{Width}\end{tabular} & \begin{tabular}[c]{@{}c@{}}\textbf{Fully} \\ \textbf{Streaming}\end{tabular} \\ 
\hline
Bitonic and odd-even sorting network (Batcher, 1968, \cite{10.1145/1468075.1468121}) & $O(n\log^2n)$ & $O(n\log^2n)$ & Flip-flop & $n$ & Yes \\
\hline
Folded bitonic sorting network (Stone, 1971, \cite{1671798}) & $O(n)$ & $O(n)$ & Flip-flop & $n$ & No \\
\hline
Odd-even transposition sorting network (Knuth, 1968, \cite{knuth1997art}) & $O(n^2)$ & $O(n^2)$ & Flip-flop & $n$ & Yes \\
\hline
Folded odd-even transposition sorting network (Knuth, 1968, \cite{knuth1997art}) & $O(n)$ & $O(n)$ & Flip-flop & $n$ & No \\
\hline
AKS sorting network (Ajtai \textit{et al.}, 1983, \cite{10.1145/800061.808726}) & $O(n\log n)$ & $O(n\log n)$ & Flip-flop & $n$ & Yes \\
\hline
Linear sorter (Lee and Tsai, 1995, \cite{Lee1995} \& Perez-Andrade \textit{et al.}, 2009, \cite{PEREZANDRADE20091705}) & $O(n)$ & $O(n)$ & Flip-flop & 1 & Yes \\
\hline
Interleaved linear sorter (ILS) (Ortiz and Andrews, 2010, \cite{5470730}) & $O(wn)$ & $O(wn)$ & Flip-flop & $1 \leq w \leq n$ & Yes \\
\hline
Shear-sort (2D mesh) (Scherson and Sen, 1989, \cite{16500}) & $O(n)$ & $O(n)$ & Flip-flop & $n$ & No \\
\hline
Streaming sorting network (Zuluaga \textit{et al.}, 2016, \cite{10.1145/2854150}) & $O(w\log^2n)$ & $O(n\log^2n)$ & RAM & $2 \leq w < n$ & Yes \\
\hline
Folded streaming sorting network (Zuluaga \textit{et al.}, 2016, \cite{10.1145/2854150}) & $O(w)$ & $O(n)$ & RAM & $2 \leq w < n$ & No \\
\hline
\end{tabular}
%\vspace{-1.5em}
\label{o_n_complexity}
%\vspace{-0.8em}
\justify{{Chen and Prasanna~\cite{Chen2017Computer} also %examine and 
assess the performance %comparison 
of sorting architectures, considering their $O(n)$ complexities. Their evaluation encompasses considerations of latency, logic usage, memory usage, throughput, and memory throughput. They study % study explores discussions concerning 
merge sort, in-place sort, bitonic sort, and sort-merge join.}}
\end{table*}

Lenjani \textit{et al.}~\cite{lenjani2022pulley} proposed \textit{Pulley}, an algorithm/hardware
co-optimization technique for in-memory sorting. Pulley %which is a first in-memory-layer approach for multi-gigabyte sorting that 
uses 3D-stacked memories. They employ Fulcrum~\cite{lenjani2020fulcrum} for the baseline PIM architecture. Fulcrum inputs data into a %the 
single-word arithmetic logic unit (ALU) in a sequential manner and enables operations that involve data dependencies as well as operations based on a predicate.
%\Amir{Lets have same pattern for abbreviations}
In Fulcrum, every pair of subarrays %also 
has three row-wide buffers called \textit{Walkers}. In the radix sorting proposed in Fulcrum, all buckets have the same length, and a bucket in each pass can always fit in one subarray. For efficient sorting %processing 
of large data %sorting
using Fulcrum, Lenjani~\textit{et~al.} modified the design by calculating the exact length of each bucket and the %exact 
position of each key within that %each
bucket.
In the first step, the keys of each processing unit are sorted locally. In this step, %In the local sorting step,
the keys are dichotomized into two buckets (\textit{Bucket0} and \textit{Bucket1}). The subarray-level processing unit (SPU) starts \textit{Bucket0} from the bottom of the space and fills it upward, and starts \textit{Bucket1} from the end of the space and fills it downward.
%\hassan{Is this correct? Bucket0 from the ``bottom'' and Bucket1 from the ``end''? are these the same?} \blue{Newsha:Yes it is correct but bottom and end are not synonymous here. bucket0(Bn-1, Bn-2, Bn-3,,,) it is filled starting from Bn-1 which is the end of the space and then it is filled downward. Bucket1(..., B3,B2,B1,B0) it is filled from B0 which is the bottom of the space and then it is filled upward to B1,B2,,,Bn.}
In the next step, each SPU generates the histogram values of the first 256 buckets iteratively, and all SPUs reduce the histogram values of each of the 256 buckets in the lowest sub-array. In Pulley, 
%\hassan{What is ``Pulley'' here?}\blue{explaination added}
each vault’s core in the logic layer performs a prefix-sum on all the shared sub-arrays in the vaults. Then, the cores in the vaults %also 
aggregate their prefix-sum arrays. They evaluate Pulley in 1-device and 6-device settings, where each device has four stacks of 8-GB memories. Compared to IMC-Sort, %their design 
Pulley has a lower working frequency.

Wu and Huang~\cite{wu2015data} introduced a novel sorting technique specifically tailored to NAND flash-based storage systems, aiming to optimize performance and efficiency. %Several key contributions are presented to achieve this goal. First, 
They propose a record rearrangement and replacement method for unclustered sorting, which involves scanning sorted tags to efficiently rearrange records and minimize unnecessary page reads during the process. They introduce
%Furthermore, 
a strategic decision rule to harness the advantages of both clustered and unclustered sorting approaches. This rule categorizes records based on their length and then selects the most appropriate sorting method (clustered or unclustered) for each category, followed by merging the sorted results. They reuse data to reduce %The concept of reused data is also presented, reducing 
page writes by detecting content similarities in the output buffer and marking logical addresses in the address translation table for potential reuse.
They provide a comprehensive I/O analysis, comparing the performance of clustered sorting, unclustered sorting, MinSort, and FAST in terms of page reads and writes. %These insights inform the development of the decision rule. 
Finally, they implement and test the proposed methods %are implemented and tested 
on real hardware, including an Intel SSD and a Hitachi HDD, demonstrating significant performance improvements compared to traditional external sorting methods. %This research presents valuable advancements in optimizing NAND flash-based storage systems, with practical applications in real-world environments.

Samardzic~\textit{et al.}~\cite{samardzic2020bonsai}  introduced ``Bonsai," an %innovative 
adaptive sorting solution that leverages merge tree architecture to optimize sorting performance across a wide range of data sizes, from megabytes to terabytes, on a single CPU-FPGA server node. Bonsai's adaptability is achieved by considering various factors, including computational resources, memory sizes, memory bandwidths, and record width. It employs analytical performance and resource models to configure the merge tree architecture to match the available hardware and problem sizes.
Their approach can enhance sorting efficiency on a single FPGA while also being used as a foundation for potential use in larger distributed sorting systems. Bonsai's primary objective is to minimize sorting time by selecting the optimal adaptive merge tree configuration based on the hardware, merger architecture, and input size. They %The article 
demonstrate the feasibility of implementing merge trees on FPGAs, highlighting their superior performance across various problem sizes, particularly for %focusing on 
DRAM-scale sorting. % on AWS EC2 F1 instances.
Bonsai achieves significant speedup over CPU, FPGA, and GPU-based sorting implementations, along with impressive bandwidth efficiency improvements, making it an appealing solution for % valuable contribution to the field of 
adaptive sorting.

\section{Open challenges}
\label{Sec:Ochall}

\color{red}

\color{black}
Although significant strides have been made in the field of hardware sorting, numerous challenges persist, warranting further research and innovation. 
%Despite the progress made thus far, hardware-based sorting still faces unresolved issues and inefficiencies. 
In this section, we explore the ongoing challenges within the research on hardware-assisted sorting. %,designs, %the field of sorting research,
%considering the previous discussions. 
Addressing these challenges can result in sorting solutions that are more efficient in different aspects, from performance to footprint area, power, and energy consumption. %from performance to both more efficient and more mindful of power consumption. 
These %specific open
challenges are
%highlighted in Fig.~\ref{challenges} and %further 
elaborated on in the following sections.

%\vspace{-0.5em}

\subsection{Algorithmic Considerations}
With recent research opportunities and emerging %in novel 
sorting solutions %paradigms like 
such as in-memory %computing sorting 
and partial sorting, future research needs to explore potential avenues for radically novel %future
sorting architectures, %spanning 
from algorithmic considerations to hardware-level enhancements. For instance, when developing new sorting algorithms, it is crucial to commence with an initial argument considering a time complexity of $O(n)$. %, as is customary.
Table~\ref{o_n_complexity} enumerates various sorting network architectures and highlights key features emphasized by Zuluaga \textit{et al.}~\cite{10.1145/2854150}. Assessing the evolution of %As we assess the evolution of chronological 
sorting architectures, an emerging trend involves %a recent trend has emerged involving 
%the utilization of 
using RAM devices for a new sorting approach known as \textit{stream sorting}~\cite{9569153, 10.1145/2854150}. %In %their study, Zuluaga \textit{et al.} investigate 
Stream sorting %, wherein the algorithm 
takes $n$ data words as input and produces $w$ words per clock cycle across $n/w$ clock cycles. The sorter achieves a throughput of $w$ if it operates in a fully streaming manner, implying no waiting time between consecutive input sets. Without a fully stream network, the throughput will be less than $w$ words per cycle.
We anticipate that one of the pivotal challenges lies in devising algorithms tailored specifically for hardware design, addressing pipeline and parallel processing concerns. Solutions such as stream sorting represent cutting-edge approaches for achieving a more efficient design right from the initial stages, optimizing both memory utilization and time complexity.

\subsection{Power and Energy Efficiency}
 The issue of power usage holds significant importance in current and future hardware designs. Given that sorting designs %networks 
 are being incorporated into a range of embedded and power-limited systems, the reduction of power consumption takes on a vital role. Future works %Researchers 
 must delve %It is imperative for researchers to delve 
 into innovative strategies for ultra-low-power hardware. These could encompass advanced clock gating, dynamic voltage scaling, and enhanced management of data transfer to curtail the energy consumption tied to the implementation of sorting designs. %networks.
 Additionally, by loosening %precision 
 accuracy demands and taking advantage of approximate computing techniques, hardware has the capacity to execute computations with fewer resources.

\subsection{Resource Limitations}

Hardware designs must operate within the boundaries defined by accessible resources such as registers, memory, and processing units. Striving to optimize the utilization of these resources while upholding performance is challenging, especially when dealing with intricate sorting algorithms that exhibit diverse computational demands. Lin~\textit{et~al.}~\cite{lin2017hardware} provide a trade-off between throughput and resources. UC-based solutions 
(\textit{e.g.,}~\cite{Najafi2018Low-Cost, Amir2022A, alam2022sorting}) have successfully achieved hardware sorting designs with extremely simple digital logic. However, they achieved this at the cost of an exponential increase in latency. Developing future sorting systems based on such emerging computing systems that operate on simple data representations~\cite{Najafi_TVLSI_2019, aygun2023HDCencoding, Najafi2018Low-Cost} is a promising path forward.

%\vspace{-0.75em}
\subsection{Latency vs. Throughput Trade-off}

Designing hardware sorting systems %frequently 
necessitates finding the right compromise between latency (the duration of %required for 
a single sorting operation) and throughput (the number %quantity 
of sorting operations %achievable 
completed within a specific %time span
period). Designers must achieve an optimum point %equilibrium 
based on the application expectations and hardware constraints. %precise demands of the intended application.

%\vspace{-0.75em}
\subsection{Parallelism}

Sorting algorithms %frequently 
encompass repetitive and regular %autonomous 
processes that hold the potential for improvement with parallelization and pipelining. Nonetheless, %the creation of effective 
implementing efficient parallel/pipelined hardware architectures (\textit{e.g.,}~\cite{papaphilippou2018flims}) and the oversight of data inter-dependencies can intricate these endeavors. Striking a harmonious equilibrium amidst diverse processing units while upholding synchronization and communication can pose a considerable challenge. PIM solutions hold significant promise for the highly parallel execution of future sorting architectures.

%\vspace{-0.75em}
\subsection{Adaptation}

Numerous practical applications demand data sorting %the sorting of data
in dynamic and ever-evolving streams. Crafting hardware-based sorting designs %networks 
capable of adeptly managing these dynamic inputs in real-time presents a multifaceted difficulty. It is imperative for researchers to delve into adaptive algorithms capable of flexibly adapting to shifting input patterns. This adaptability should ensure sustained, efficient sorting performance while minimizing any notable additional workload.

%\vspace{-0.5em}
\subsection{Customization}

Hardware sorting designs may need to be customized for specific applications or environments. This requires flexibility in the design process (\textit{e.g.,}~\cite{lin2017hardware, papaphilippou2018flims}) to accommodate different requirements. From different data types to various data precisions (\textit{i.e.,} bit-widths), size of the dataset, and hardware constraints (\textit{e.g.,} area and power budget), achieving the best performance may require customized hardware. However, the higher design time and cost of implementing customized hardware must also be considered.

%\vspace{-0.5em}
\subsection{Data Movement and Memory Access}

 Optimal memory access is pivotal for sorting algorithms, and hardware architectures must strive to curtail data transfer and cache-related inefficiencies. Sorting entails frequent data comparisons and exchanges, introducing the potential for irregular memory access patterns. Effectively handling these access patterns is imperative to avert potential performance bottlenecks. The problem aggregates in big data applications where the sorting engine is expected to sort a large set of data.

%\vspace{-0.5em}
\subsection{Technology Scaling}

 Hardware designs might necessitate adjustments to accommodate technological shifts, such as advancements in the semiconductor manufacturing process. %techniques.
 Designers must meticulously evaluate the potentials and consequences of technological scaling on factors such as performance, area, power and energy usage, and various design parameters.

\ignore{

\subsection{In memory sorting}

In our future works, we envision a comprehensive expansion of the innovative architectures we have introduced. Our roadmap includes broadening the scope of these architectures to encompass a variety of sorting applications.
 For instance, we aim to develop highly efficient in-memory implementations of image and signal processing applications. 
 we also plan to uncover novel ways in which In-memory sorting can revolutionize data transmission, compression, and error correction processes in the communications and coding domain applications.
}

%\vspace{-0.5em}

\section{Conclusion}

\label{Sec:Conclusion} 

Sorting is one of the crucial operations in computer science, widely used in many application domains, %and has been used in a wide range of applications, 
from data merging to big data processing, database operations, robotics, wireless sensor networks, signal processing, and wireless networks. A substantial body of work is dedicated to designing hardware-based sorting. %customized sorting hardware. 
In this survey, we reviewed the latest developments in hardware-based sorting, encompassing both comparison-based and comparison-free solutions. Comparison-based solutions tend to incur high hardware costs, particularly as the volume and precision of data increase. Comparison-free solutions have recently been proposed to overcome the challenges associated with compare-and-swap-based sorting designs. %, especially as the number and precision of input data increase.
We reviewed recent hardware solutions for partial sorting and stream sorting, which are used to sort the top-$k$ largest or smallest values of the dataset. % out of $N$ elements, where $k<N$. 
We also studied the latest emerging in-memory solutions for sorting operations. Finally, we outlined the %significant open 
 challenges in developing future hardware sorting, aiming to provide readers with insights into the next generation of sorting systems. %future sorting scheme design.
 
%\lipsum[1-2]

\color{black}

%\section*{References}

%{%\normalsize
%\renewcommand{\bibfont}{\huge}
%\huge
%\bibliographystyle{abbrv}
%\IEEEtriggeratref{1}
%\vspace{-0.5em}

\bibliographystyle{ieeetr}
\bibliography{References}

\balance

\begin{IEEEbiography}[{\includegraphics[width=1in,height=1.25in,clip,keepaspectratio]{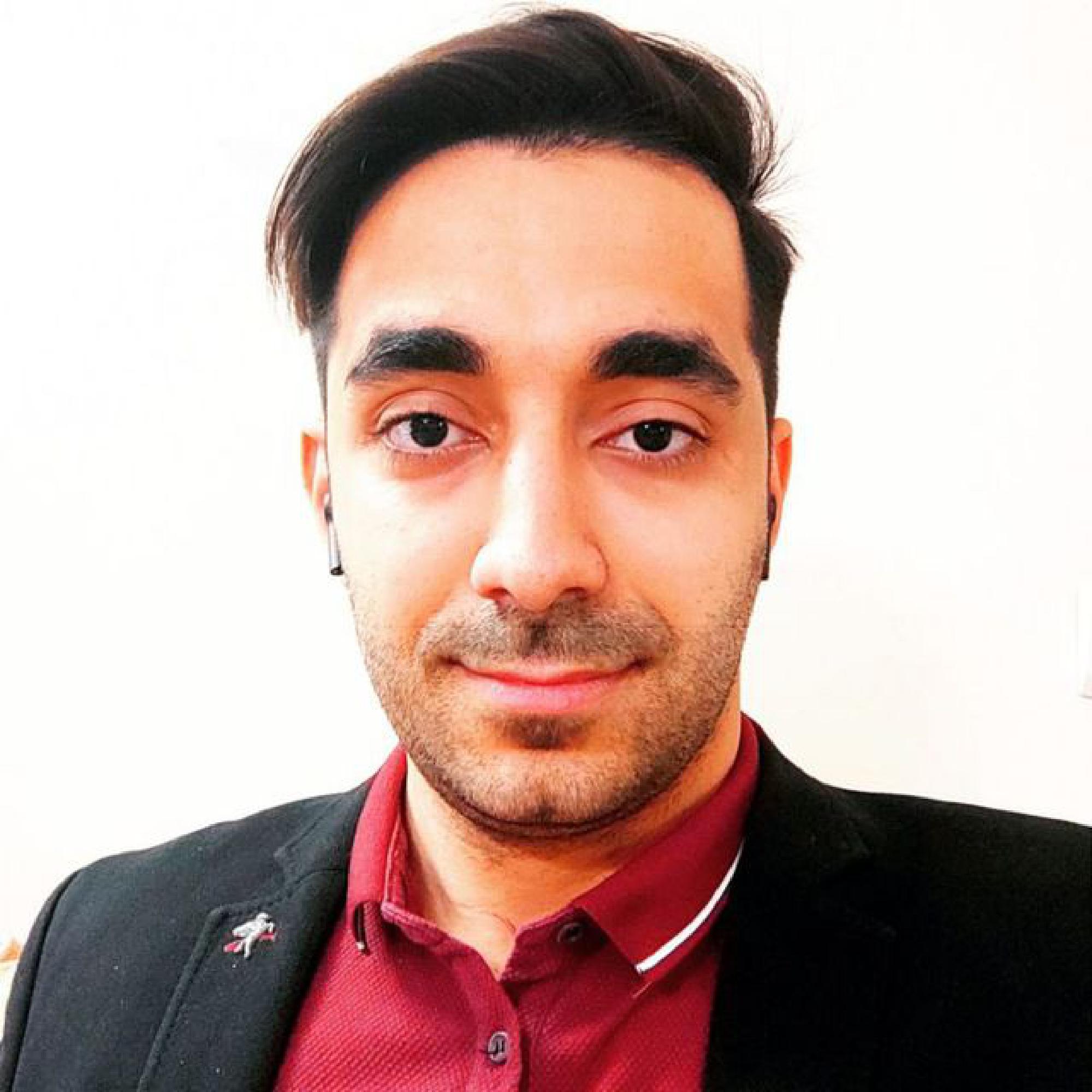}}]{Amir Hossein Jalilvand}
received a B.Sc. degree in Computer Engineering from Bu-Ali Sina University, Hamadan, Iran, and an M.Sc. degree in Computer Engineering - Computer Architecture from Iran University of Science and Technology, Tehran, Iran. Currently, he is a Ph.D. candidate in Computer Engineering. His research interests include Stochastic and Unary Computing, Computer Architecture,  Fuzzy Logic, and machine learning.
\end{IEEEbiography}

\vspace{-5pt}

\begin{IEEEbiography}[{\includegraphics[trim={0.12in 0in 0.12in 0in},width=1in,height=1.25in,clip,keepaspectratio]{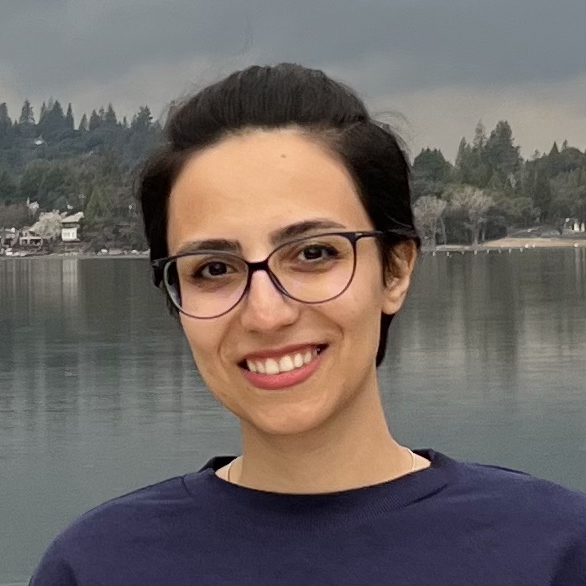}}]{Faeze S. Banitaba}
(S’23) received a B.Sc. degree in Computer Engineering from Shahid Beheshti University, Tehran, Iran. She completed her M.Sc. degree in Computer Architecture from the University of Tehran in 2017. Currently, she is a Ph.D. student in Computer Engineering. Her research interests rely on Stochastic Computing, Hardware Security for Neural Networks, and Hyperdimensional Computing. She was recognized as a DAC Young Fellow 2023. 
\end{IEEEbiography}

\vspace{-5pt}

\begin{IEEEbiography}[{\includegraphics[width=1in,height=1.25in,clip,keepaspectratio]{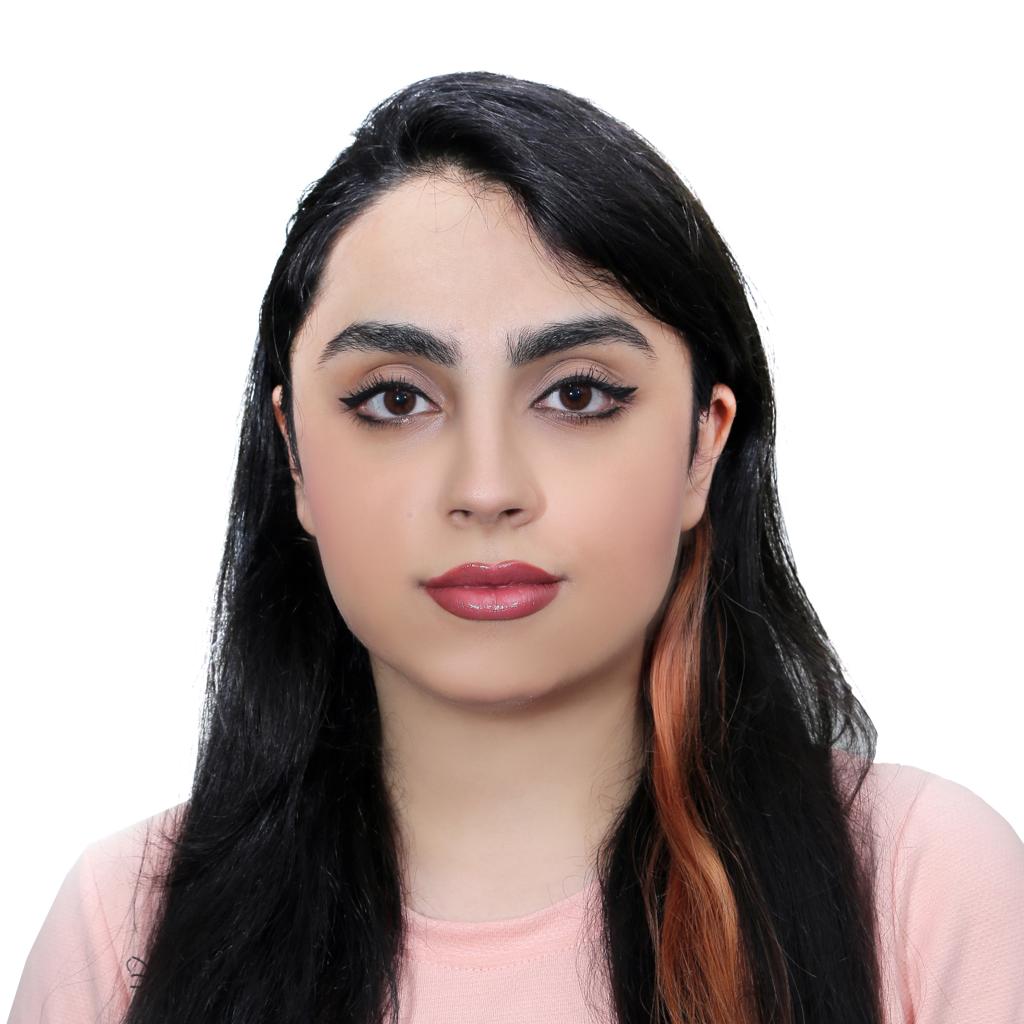}}]{Seyedeh Newsha Estiri}
received a B.Sc. degree in Computer Engineering - Hardware System Design from Iran University of Science and Technology, Tehran, Iran, in 2020. Currently, she is a second-year Ph.D. student in computer engineering at the University of Louisiana at Lafayette. Her research interests include Stochastic computing,  Machine Learning Applications and Hardware, and Computer Architecture. Newsha was selected as a DAC Young Fellow in DAC 2022.
\end{IEEEbiography}

\vspace{-5pt}

\begin{IEEEbiography}[{\includegraphics[width=1in,height=1.15in,clip,keepaspectratio]{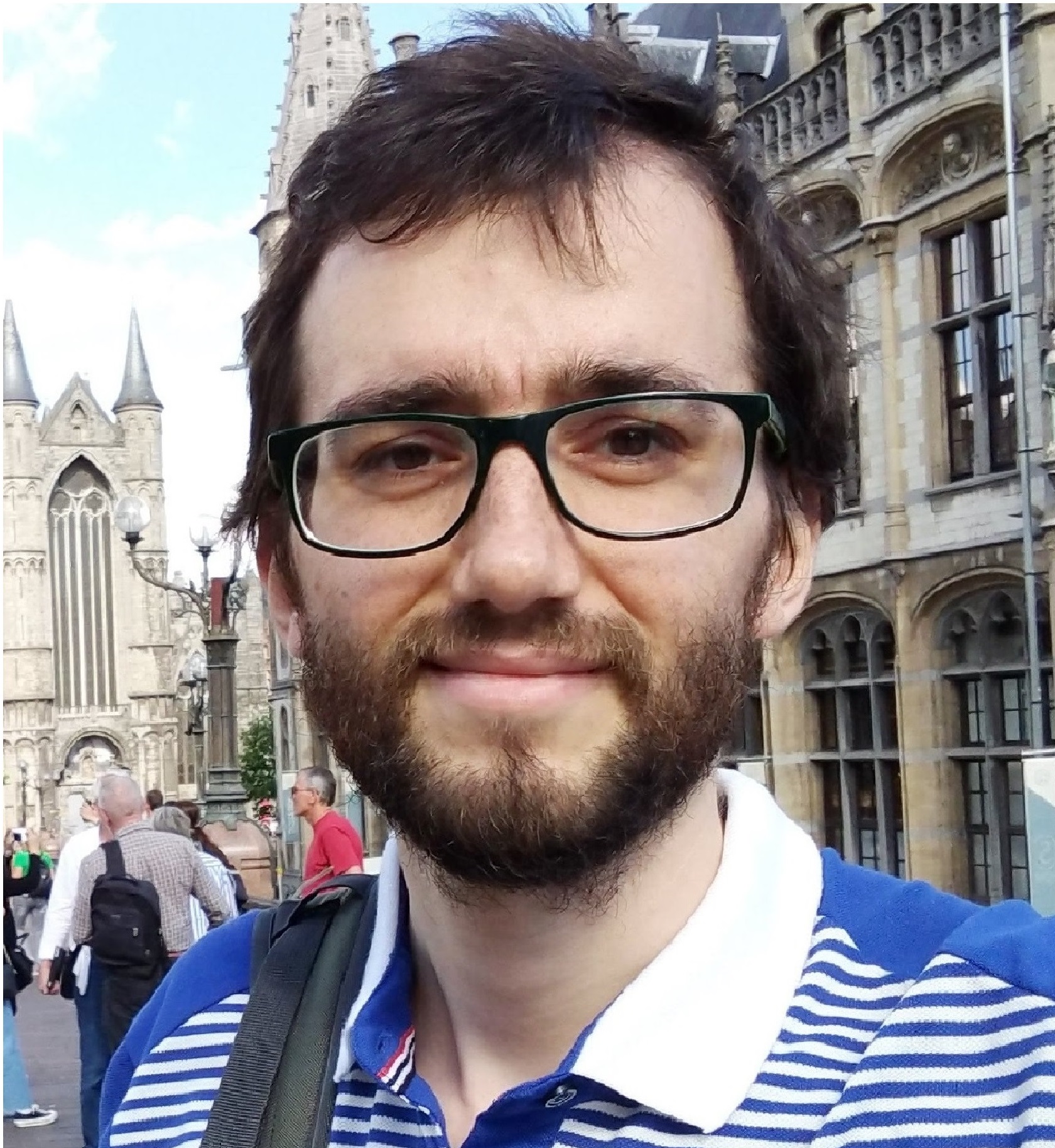}}]{Sercan Aygun}
(S’09-M’22) received a B.Sc. degree in Electrical \& Electronics Engineering and a double major in Computer Engineering from Eskisehir Osmangazi University, Turkey, in 2013. He completed his M.Sc. degree in Electronics Engineering from Istanbul Technical University in 2015 and a second M.Sc. degree in Computer Engineering from Anadolu University in 2016. Dr. Aygun received his Ph.D. in Electronics Engineering from Istanbul Technical University in 2022. Dr. Aygun’s Ph.D. work has appeared in several Ph.D. Forums of top-tier conferences, such as DAC, DATE, and ESWEEK. He received the Best Scientific Research Award of the ACM SIGBED Student Research Competition (SRC) ESWEEK 2022 and the Best Paper Award at GLSVLSI'23. Dr. Aygun's Ph.D. work was recognized with the Best Scientific Application Ph.D. Award by the Turkish Electronic Manufacturers Association. He is currently a postdoctoral researcher at the University of Louisiana at Lafayette, USA. He works on emerging computing technologies, including stochastic computing in computer vision and machine learning.
\end{IEEEbiography}

\vspace{-5pt}

\begin{IEEEbiography}[{\includegraphics[width=1in,height=1.17in,clip,keepaspectratio]{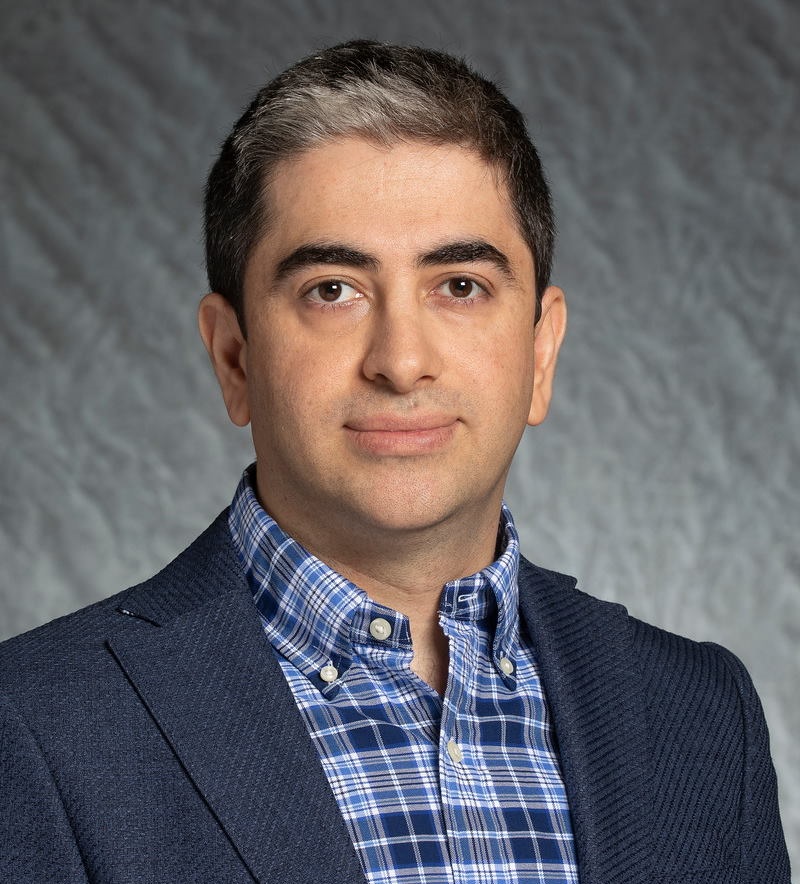}}]{M. Hassan Najafi}
(S’15-M’18-SM'23) received the B.Sc. degree in Computer Engineering from the University of Isfahan, Iran, the M.Sc. degree in Computer Architecture from the University of Tehran, Iran, and the Ph.D. degree in Electrical Engineering from the University of Minnesota, Twin Cities, USA, in 2011, 2014, and 2018, respectively. He is currently an Assistant Professor with the School of Computing and Informatics, University of Louisiana, LA, USA. His research interests include stochastic and approximate computing, unary processing, in-memory computing, and hyperdimensional computing. He has authored/co-authored more than 60 peer-reviewed papers and has been granted 5 U.S. patents with more pending. In recognition of his research, he received the 2018 EDAA Outstanding Dissertation Award, the Doctoral Dissertation Fellowship from the University of Minnesota, and the Best Paper Award at the ICCD’17 and GLSVLSI'23. Dr. Najafi has been an editor for the IEEE Journal on Emerging and Selected Topics in Circuits and Systems.
\end{IEEEbiography}

\end{document}